\documentclass[twocolumn,twocolappendix,,tighten]{aastex701}

\usepackage{tikz}
\usepackage{array}
\usepackage{multirow}
\usepackage{float}
\usepackage{subcaption}
\usepackage[utf8]{inputenc}
\usepackage{graphicx,color}
\usepackage{mathtools}	
\usepackage{amsmath}
\usepackage{amssymb}
\usepackage{bm}
\usepackage{pstricks} 
\usepackage[normalem]{ulem}
   \usepackage[british]{babel}             
   \usepackage{newtxtext}                  
   \usepackage[slantedGreek]{newtxmath}    
   \usepackage[T1]{fontenc}                
\def\aap{A\&A}
\def\apjl{ApJ}

\def\apjs{ApJS}

\newcommand{\f}{_\textrm{0}}

\newcommand{\D}{_\mathrm{D}}						
\newcommand{\z}{_\mathrm{0}}					   	
\newcommand{\turb}{_\mathrm{t}}			   		

\renewcommand{\theenumi}{\arabic{enumi})}

\newcommand{\A}{_\mathrm{A}}
\newcommand{\I}{_\mathrm{I}}

\newcommand{\soft}{_\mathrm{soft}}

\newcommand{\jet}{_\mathrm{j}}

\newcommand{\acc}{_\mathrm{acc}}
\newcommand{\Edd}{_\mathrm{Edd}}

\newcommand{\Rsun}{\,\mathrm{R_\odot}}
\newcommand{\Msun}{\,\mathrm{M_\odot}}
\newcommand{\Msunyr}{\,\mathrm{M_\odot\,yr^{-1}}}

\newcommand{\bind}{_\mathrm{bind}}

\newcommand{\sound}{_\mathrm{s}}

\newcommand{\half}{_\mathrm{h}}

\newcommand{\restart}{_\mathrm{res}}

\newcommand{\luke}[1]{\textcolor{red}{#1}}

\newcommand{\erg}{\,{\rm erg}}
\newcommand{\g}{\,{\rm g}}
\newcommand{\gcmcmcm}{\,{\rm g\,cm^{-3}}}

\newcommand{\km}{\,{\rm km}}
\newcommand{\kms}{\,{\rm km\,s^{-1}}}

\newcommand{\K}{\,{\rm K}}

\newcommand{\p}{\,{\rm pc}}

\newcommand{\s}{\,{\rm s}}
\newcommand{\yr}{\,{\rm yr}}     
\newcommand{\da}{\,{\rm d}}     
\newcommand{\dyncmcm}{\,{\rm dyn\,cm^{-2}}}

\begin{document}

\title[Effect of NS Jets on CE Evolution]{Effect of Neutron Star Jets on Common Envelope Evolution}

\author[sname='Gurjar']{Deepanshu Gurjar}
\affiliation{National Institute of Science Education and Research, Bhubaneswar 752050, India}
\affiliation{Homi Bhabha National Institute, Training School Complex, Anushakti Nagar, Mumbai 40094, India}
\email{deepanshugurjar99@gmail.com}  

\author[orcid=0000-0003-4935-5550,sname='Chamandy']{Luke Chamandy}
\affiliation{National Institute of Science Education and Research, Bhubaneswar 752050, India}
\affiliation{Homi Bhabha National Institute, Training School Complex, Anushakti Nagar, Mumbai 40094, India}
\email[show]{lchamandy@niser.ac.in}  
\correspondingauthor{Luke Chamandy}

\author[sname='Blackman']{Eric G. Blackman}
\affiliation{Department of Physics and Astronomy, University of Rochester, Rochester NY 14627, USA}
\email{blackman@pas.rochester.edu}

\author[sname='Zou']{Yangyuxin Zou}
\affiliation{Department of Physics and Astronomy, University of Rochester, Rochester NY 14627, USA}
\email{yzou5@ur.rochester.edu}

\author[sname='Liu']{Baowei Liu}
\affiliation{Center for Integrated Research Computing, University of Rochester, Rochester NY 14627, USA}
\affiliation{Department of Physics and Astronomy, University of Rochester, Rochester NY 14627, USA}
\email{baowei.liu@rochester.edu}

\author[sname='Nordhaus']{Jason Nordhaus}
\affiliation{National Technical Institute for the Deaf, Rochester Institute of Technology, Rochester, NY 14623, USA}
\affiliation{Center for Computational Relativity and Gravitation, Rochester Institute of Technology, Rochester, NY 14623, USA}
\email{jtnsma@rit.edu}

\defcitealias{Zou+22}{Z22}

\begin{abstract}
The common envelope (CE) phase is a key stage in binary star evolution that is still not very well understood.  Once engulfed by the giant star, the binary companion may accrete envelope material.  For neutron star (NS) companions, such accretion may in principle occur at mass rates several orders of magnitude above the Eddington limit and may result in outflows dominated by powerful bi-polar jets with mass-loss rates similar to the accretion rates.  Such jets would impact the morphology of the system and the rate of envelope unbinding, which affect the duration and outcome of the CE event.  Employing 3D global hydrodynamic simulations, we study the role of such NS jets in a CE event involving a red giant branch star.  The jets eventually drill through and break out of the envelope, producing prominent low-density bi-polar lobes.  The jets cause about twice as much envelope mass to be unbound ($\sim20\%$ of the envelope) as compared to simulations of the same duration without NS jets ($\sim10\%$).  However, the rate of mass unbinding due to the jets decreases towards the ends of the simulations as the jets break out and energetically decouple from the envelope.  Moreover, jet activity leads to slightly reduced drag on the binary, decreasing the rate of orbital energy transfer to the envelope.  Hence, while such powerful jets can play an important role, negative feedback effects tend to prevent them from dominating envelope unbinding and dictating CE outcomes.
\end{abstract}

\keywords{\uat{Common envelope evolution}{2154}; \uat{Common envelope binary stars}{2156}; \uat{Interacting binary stars} {801}; \uat{Neutron stars}{1108}; \uat{Stellar jets}{1607}; \uat{Stellar mass loss}{1613}; \uat{Stellar accretion}{1578}}

\section{Introduction}
A binary stellar system undergoing mass transfer may become unstable and enter the common envelope (CE) phase \citep{Paczynski76},
wherein the core of the giant (donor) star and more compact main sequence (MS), 
sub-stellar or stellar remnant companion inspiral inside a shared atmosphere or envelope.
If enough orbital energy and angular momentum are transferred to the envelope during the CE event,
it can be ejected, as evidenced, for instance, by the existence of numerous small-period ``post-CE'' binaries,
as well as by some long-duration simulations \citep[e.g.,][]{Ondratschek+22}.
However, a longstanding question is to what extent other sources of energy may assist in unbinding the envelope
(for reviews see \citealt{Ivanova+13a}, \citealt{Ivanova+20} and \citealt{Roepke+Demarco22}).
One possible source of energy is that released 
when gas in the envelope accretes onto the companion.
During the CE phase, 
envelope material falls toward the companion at super-Eddington rates 
\citep{Armitage+Livio00,Ricker+Taam08,Ricker+Taam12,Macleod+Ramirez-ruiz15a,Chamandy+18}.
As the envelope is optically thick, 
the accretion flow could result in a buildup of pressure which rapidly quenches it \citep{Chamandy+18}. 
However, bipolar jets may in principle arise and act as a pressure valve, 
directing some (even most) of the material in the accretion flow back into the envelope,
energizing it and possibly helping to unbind it \citep[e.g.,][]{Soker17b}.
If the base of the accretion flow is hot enough, 
neutrinos would form and carry away much of the energy, 
allowing accretion rates to exceed Eddington by several orders of magnitude and potentially also causing powerful jets \citep[e.g.,][]{Cheong+26}.
As well as possibly playing a role in envelope unbinding, 
such bipolar jets could affect the morphology of the CE and various observations during and after the CE event
\citep[e.g.,][]{Schreier+23,Hillel+23}.

Recent simulation studies of CE evolution that include companion jets find that
MS and even white dwarf (WD) jets are choked inside the envelope 
(\citealt{Lopez-camara+22}, \citealt{Zou+22}, hereafter \citetalias{Zou+22}).
As long as the jets are choked, 
they deposit energy into the envelope without significantly affecting the morphology of the CE.
\citetalias{Zou+22} found that jets from MS companions could increase the rate of unbinding
of envelope mass by $\sim1\%$, but that this can rise to $\sim10\%$ for WD jets.
This additional fractional unbound mass caused by the jets is comparable 
to the ratio of injected jet energy to the orbital energy released during the inspiral.

A neutron star (NS) jet would have a much higher speed, comparable to the escape speed near the NS surface.
In \citetalias{Zou+22}, 
the mass outflow rate of the WD jet $\dot{M}\jet$ was assumed to be equal to the Eddington rate $\dot{M}\Edd$.
This is plausible because while perhaps only a fraction (say $0.1$) of the accreting mass is deposited in the jet,
this can perhaps be compensated by super-Eddington accretion 
facilitated by the non-spherical flow geometry \citep[e.g.,][]{Frank+02}.
Since $\dot{M}\Edd$ is roughly proportional to the stellar radius $R$ \citep[e.g.,][]{Chamandy+18},
and the escape speed is proportional to $R^{-1/2}$,
we  might expect the jet power for a NS to be roughly the same as that of a WD or MS star if both accrete at their respective Eddington rates.
However, NS accretion may not be subject to the Eddington limit for photons 
because the emission may be dominated by neutrinos when photons are advected in the accretion flow \citep{Houck+Chevalier91,Chevalier93}.
In any case, it is possible that a large fraction of the accreting material is channeled into jets,
and the NS jet outflow rate may exceed the Eddington rate by several orders of magnitude in some cases \citep{Armitage+Livio00}. 
The jet power would be very high compared to MS or WD jets and this scenario is the focus of the present work.
We emphasize that the existence of highly super-Eddington jets, while plausible, is not yet supported by direct evidence; 
nor are the expected properties of such jets very well constrained.
Therefore, the present work is intended to be exploratory in nature.

In Section~\ref{sec:methods} we summarize the computational methods and describe the simulation runs carried out,
in Section~\ref{sec:results} we present our simulation results and in Section~\ref{sec:extrapolation} 
we discuss caveats of the model and speculate on the evolution of the system on longer timescales. 
We summarize and conclude in Section~\ref{sec:conclusions}.

\begin{deluxetable}{@{}ccccccc@{}}
\tablecaption{List of simulations performed. 
Columns show the Run ID, the maximum level of the adaptive mesh refinement (higher means better resolution), 
whether subgrid accretion is activated or not,
the restart time from Run~J8 (at which time the NS jets are activated),
the jet mass-loss rate, the characteristic jet speed 
(the speed with which the jet material would have been initialized had there been no other matter in the initialization region),
and the jet power.
The no-jet run (NJ1) and the WD jet run (J8) refer to runs from \citetalias{Zou+22}.
Note that the Eddington accretion rate can be estimated as $\dot{M}\Edd\sim2.1\times10^{-3}(R/\!\Rsun)\Msunyr$ 
\citep[e.g.,][]{Chamandy+18}, which gives $3.6\times10^{-8}\Msunyr$ for a NS of radius $12\km$.
This implies that $\dot{M}\jet\sim6\times10^4\dot{M}\Edd$ for Run~09 and $\dot{M}\jet\sim6\times10^3\dot{M}\Edd$ 
for the other NS runs.
Note that Runs~03 and 05 differ in only one inconsequential parameter, not listed (see Sec.~\ref{sec:runs} for details).
\label{tab:runs}
}
\tablehead{
Run &AMR    &Accretion    &$t\restart$ &$\dot{M}\jet$      &$v\jet$          &$\tfrac{1}{2}\dot{M}\jet v\jet^2$  \\
    &       &             &$\!\da$     &($\!\Msunyr$)      &($\!\kms$)       &($\!\erg\s^{-1}$)                  
}
\startdata
NJ1 &4      &Off          &---         &---                &---              &---                                \\
J8  &4      &On           &---         &$2\times10^{-5}$   &$8640$           &$4.7\times10^{38}$                 \\
01  &4      &On           &$19.9$      &$2\times10^{-4}$   &$8640$           &$4.7\times10^{39}$                 \\
03  &4      &On           &$19.9$      &$2\times10^{-4}$   &$3\times10^4$    &$5.7\times10^{40}$                 \\
04  &4      &Off          &$19.9$      &$2\times10^{-4}$   &$3\times10^4$    &$5.7\times10^{40}$                 \\
05  &4      &On           &$19.9$      &$2\times10^{-4}$   &$3\times10^4$    &$5.7\times10^{40}$                 \\
06  &5      &On           &$19.9$      &$2\times10^{-4}$   &$3\times10^4$    &$5.7\times10^{40}$                 \\
07  &3      &On           &$19.9$      &$2\times10^{-4}$   &$3\times10^4$    &$5.7\times10^{40}$                 \\
08  &4      &On           &$13.8$      &$2\times10^{-4}$   &$3\times10^4$    &$5.7\times10^{40}$                 \\
09  &4      &On           &$19.9$      &$2\times10^{-3}$   &$3\times10^4$    &$5.7\times10^{41}$                 \\
11  &4      &Off          &$13.8$      &$2\times10^{-4}$   &$3\times10^4$    &$5.7\times10^{40}$                 \\
\enddata
\end{deluxetable}

\begin{figure*}
  \centering
  \begin{subfigure}[b]{0.9\textwidth}
    \includegraphics[width=\textwidth]{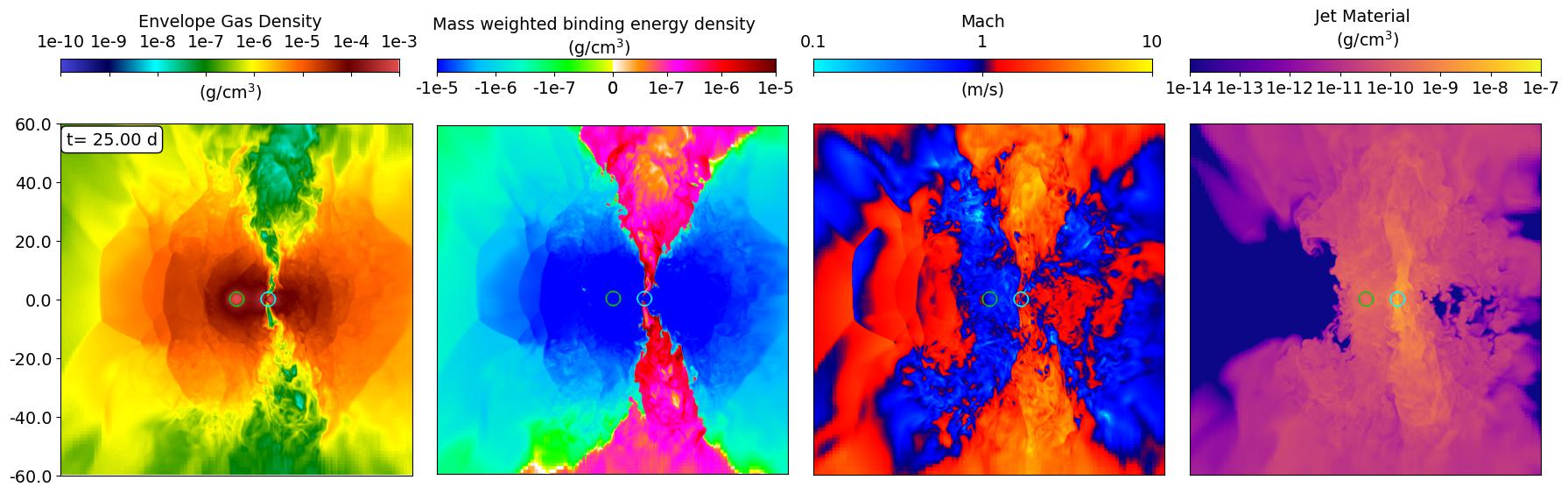}
  \end{subfigure}
  
  \begin{subfigure}[b]{0.9\textwidth}
    \includegraphics[width=0.989\textwidth]{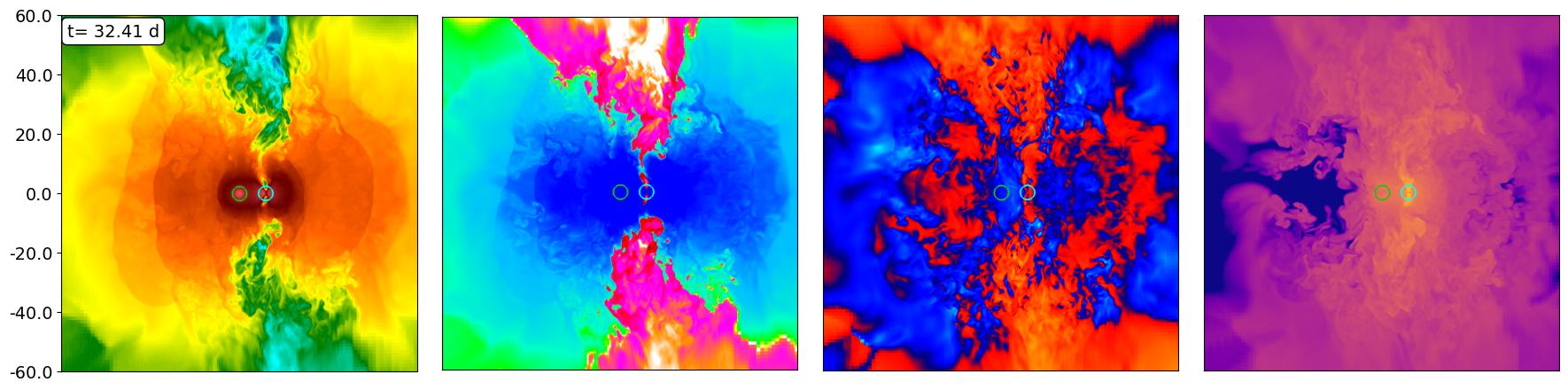}
  \end{subfigure}
  
  \begin{subfigure}[b]{0.9\textwidth}
    \includegraphics[width=0.995\textwidth]{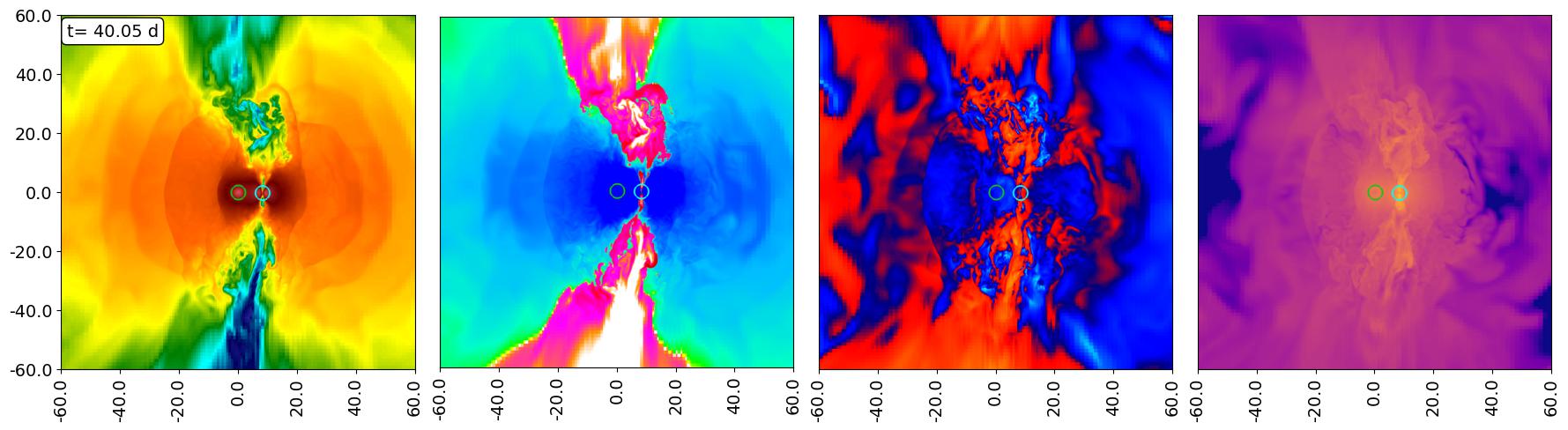}
  \end{subfigure}
  
  \caption{Snapshots of Run~05 showing zoomed-in slices through the particles of various quantities, 
    with the RGB core at the center and the NS companion to its right. 
    Softening spheres are shown with circles and axis units are $\!\Rsun$.
    The progression through time is shown from top to bottom.
    Movies are available at 
    \luke{\url{https://doi.org/10.5281/zenodo.21982012}}.
  }
  \label{fig:run05}
\end{figure*}

\begin{figure*}
  \centering
  \begin{subfigure}[b]{0.9\textwidth}
    \includegraphics[width=\textwidth]{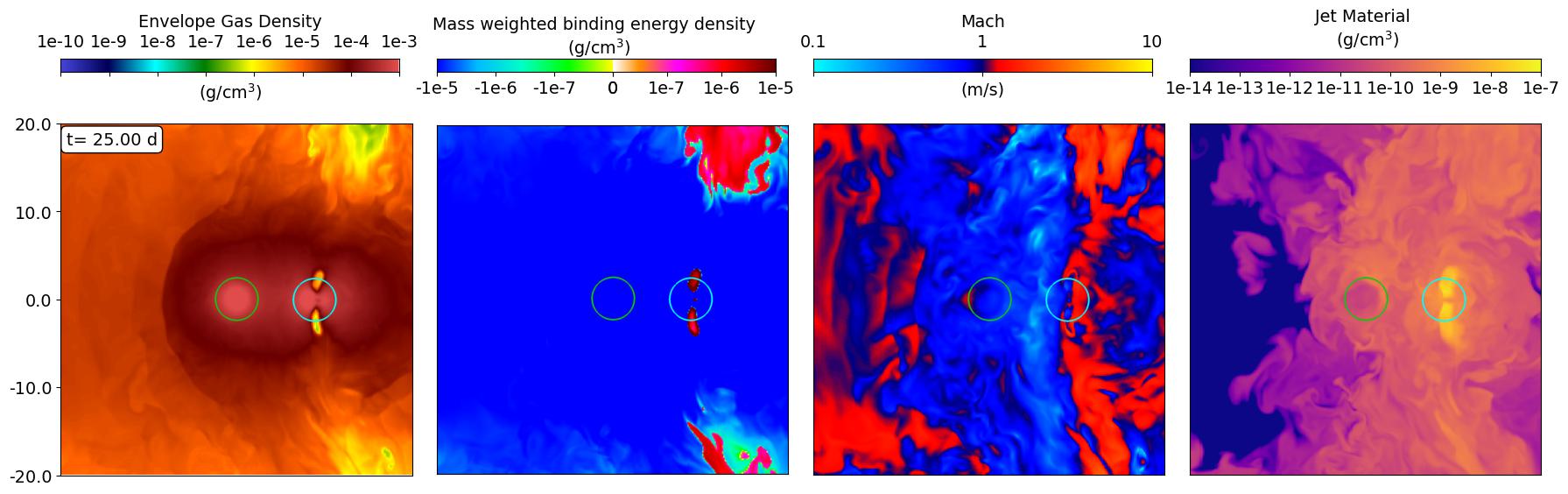}
  \end{subfigure}
  
  \begin{subfigure}[b]{0.9\textwidth}
    \includegraphics[width=0.989\textwidth]{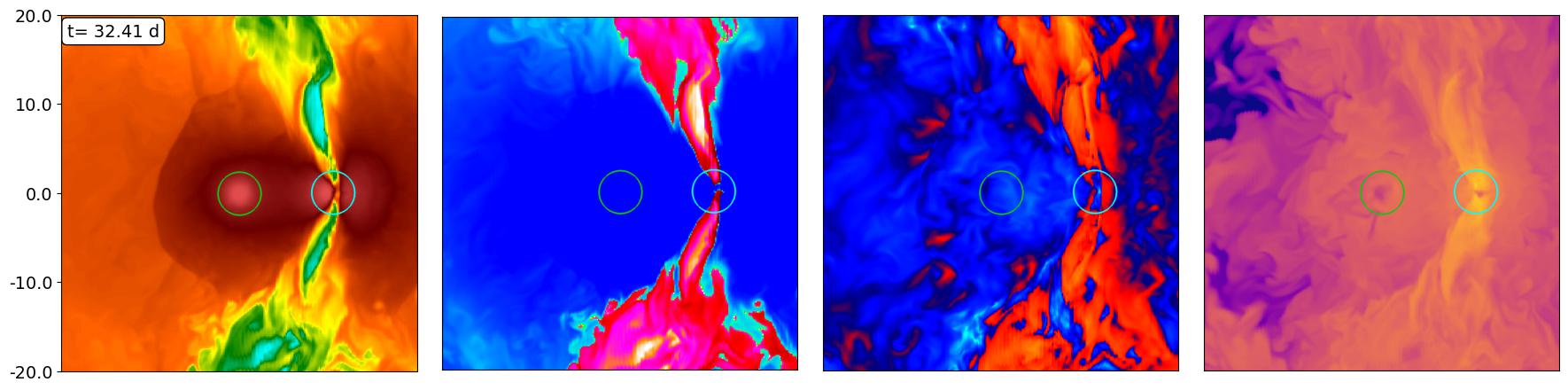}
  \end{subfigure}
  
  \begin{subfigure}[b]{0.9\textwidth}
    \includegraphics[width=0.995\textwidth]{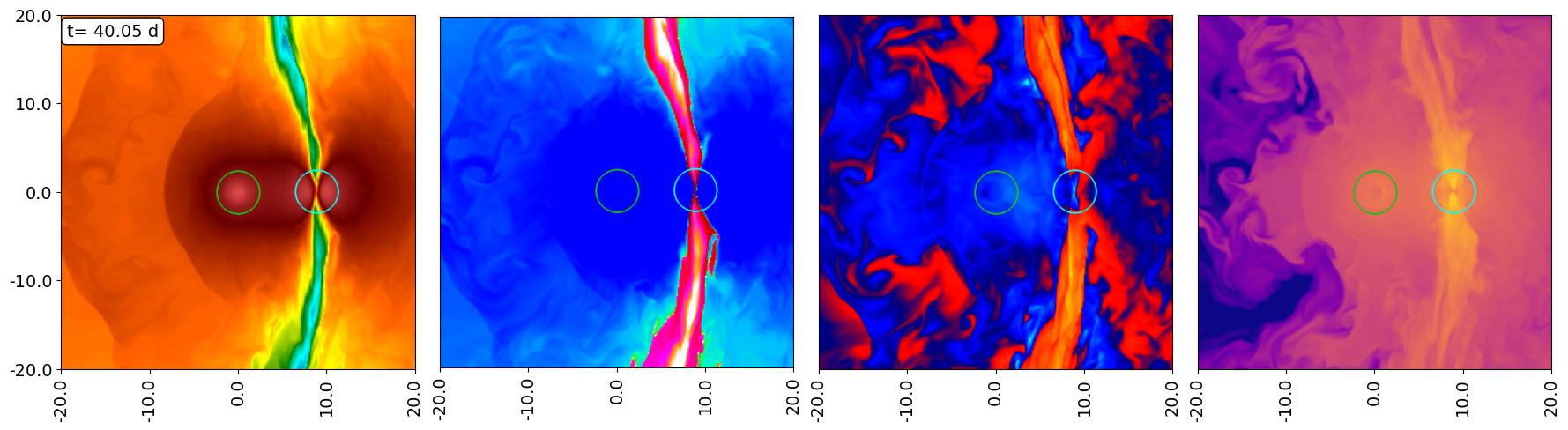}
  \end{subfigure}
  
  \caption{As Fig.~\ref{fig:run05} but now zoomed in by a factor of three.}
  \label{fig:run05_zoom}
\end{figure*}

\begin{figure*}
  \centering
  \begin{subfigure}[b]{0.9\textwidth}
    \includegraphics[width=\textwidth]{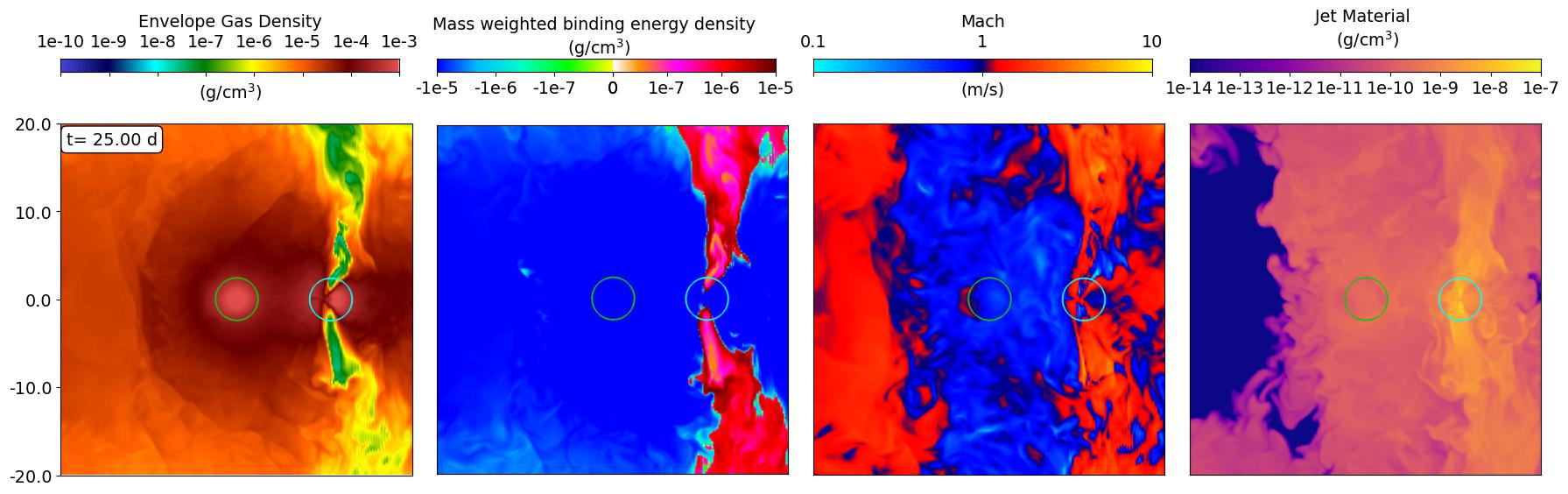}
  \end{subfigure}
  
  \begin{subfigure}[b]{0.9\textwidth}
    \includegraphics[width=0.989\textwidth]{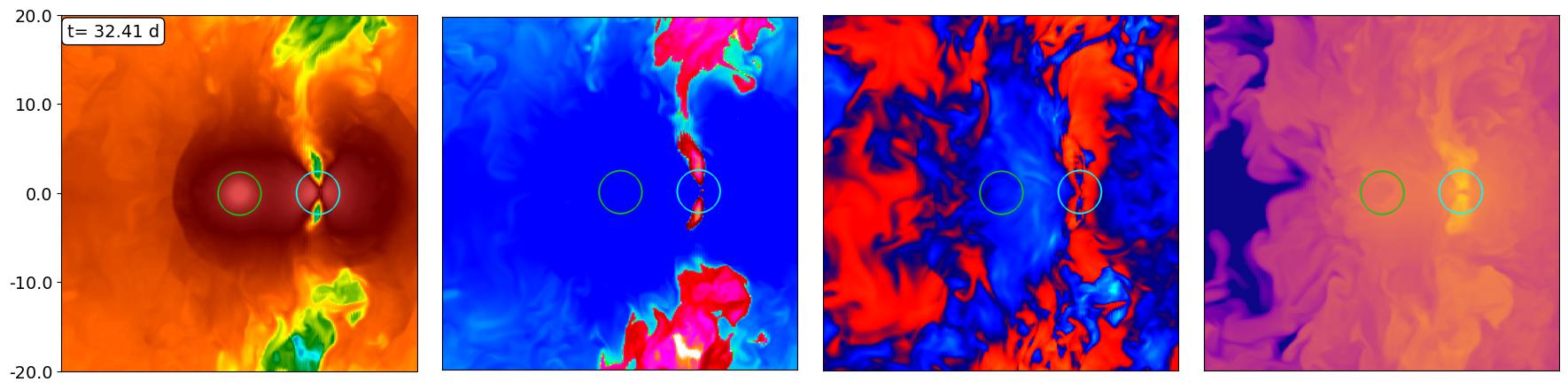}
  \end{subfigure}
  
  \begin{subfigure}[b]{0.9\textwidth}
    \includegraphics[width=0.995\textwidth]{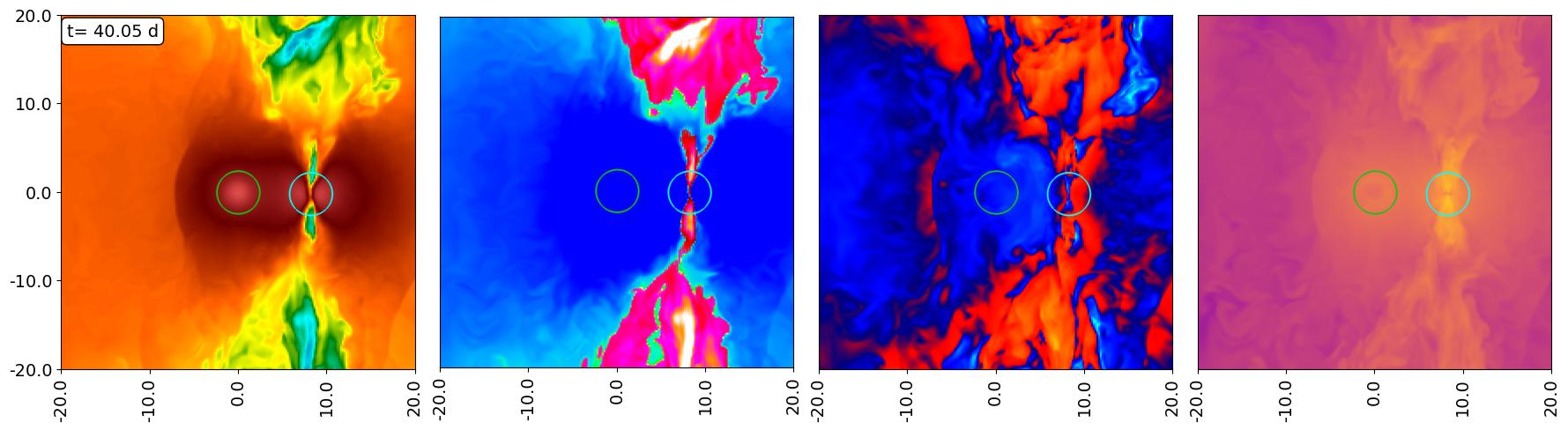}
  \end{subfigure}
  
  \caption{As Fig.~\ref{fig:run05} but now for Run~08, for which the jet is initiated at $t=13.8\da$ instead of $t=19.9\da$.}
  \label{fig:run08_zoom}
\end{figure*}

\section{Methods}\label{sec:methods}
\subsection{Common Envelope Simulation}
The binary system consists of a $2.0\Msun$, $48\Rsun$ red giant branch (RGB) star primary 
with a $0.37\Msun$ point particle representing its core and a $1.0\Msun$ point particle secondary representing a NS 
initialized in a circular orbit just outside the RGB surface.
The fact that the mass of the secondary is less than that of a typical NS 
is not expected to affect the main conclusions of this study.
The core of the primary and secondary are both gravitation-only particles with spline softening radius equal to $r\soft=2.41\Rsun$.
\citetalias{Zou+22} performed simulations with jet mass-loss rates and powers appropriate for MS and WD companions,
whereas in the present paper we explore NS companions with much higher jet powers.
In order to facilitate comparison with previous work, 
we use the same setup and parameter values as in \citetalias{Zou+22}, other than the jet mass outflow rate and power.

The system is evolved by solving the Euler equations 
using the adaptive mesh refinement (AMR) code \textsc{astrobear} \citep{Carroll-Nellenback+13},
with an ideal gas equation of state and adiabatic index $\gamma = 5/3$. 
All gravitational interactions (particle-particle, particle-gas and gas self-gravity) are fully accounted for.
The RGB profile was obtained by first evolving an MS star of zero-age mass $2\Msun$ 
with the 1D stellar evolution code \textsc{mesa} \citep{Paxton+11,Paxton+13,Paxton+15,Paxton+19},
and then replacing its core with a gravitation-only point particle and modified gas profile obtained by solving a modified Lane-Emden equation
to make it resolvable in the 3D simulations \citep{Ohlmann+17,Chamandy+18}.
The ambient medium was initialized with a uniform gas of pressure $1.0 \times 10^5\dyncmcm$
(chosen to maintain stability of the stellar surface) and a density $6.7\times 10^{-9}\gcmcmcm$ 
(equal to the surface density of the star and chosen to minimize computational cost).
As in \citetalias{Zou+22}, the ambient medium is accounted for in our analysis.
The simulation employs a box with side length $L = 1150 \Rsun$, extrapolated boundary conditions, 
and four levels of AMR with base resolution $\delta_0 = 2.25\Rsun$ and highest resolution $\delta_4 = 0.14\Rsun$, 
where $\delta_n=\delta_0/2^n$ is the cell size for AMR level $n$.
In two runs, the maximum resolution is instead set to a different value ($\delta_3=0.28\Rsun$ or $\delta_5=0.07\Rsun$)
in order to test convergence of the numerical solutions.
Simulations are conducted in a reference frame in which the center of mass remains approximately fixed.
For more details about the numerics and setup of the CE problem, we refer the reader to \citet{Chamandy+18},
\citet{Chamandy+19a}, \citetalias{Zou+22}, and \citet{Chamandy+24}. 

\subsection{Jet Prescription}
The jet outflow model is the same as in \citetalias{Zou+22} 
and uses a slightly modified version of the subgrid model from \citet{Federrath+14}; 
the reader is referred to the appendix of \citetalias{Zou+22} for the details of the numerical algorithm.
During a simulation, 
jets are fed with mass-loss rate $\dot{M}\jet$ divided equally between two spherical sectors,
which extend from the sink particle to $r\jet = \delta_0 =2.25\Rsun$ 
(inside the softening sphere of radius $2.41\Rsun$) and have half-opening angle $\theta\half$. 
The jet density radial profile peaks at $r\jet/2$ and the outflow radial velocity 
is strongly peaked for $|\theta| < \theta\half/6$, resulting in a narrow jet surrounded by a broader wind
(below we do not make this distinction and refer to the outflow as bi-polar jets).
For all of the simulations in this work, $\theta\half = 15^\circ$.
The jet subgrid model conserves mass by reducing the mass of the companion particle at the rate $-\dot{M}\jet$,
and also conserves momentum and angular momentum, as explained in \citetalias{Zou+22}.
The rate of kinetic energy supplied to the jets is approximately $\tfrac{1}{2}\dot{M}\jet v\jet^2$,
distributed across the jet initialization region as explained in \citetalias{Zou+22}.%
\footnote{In \citetalias{Zou+22} the injected jet power was erroneously stated to be $20$ times lower than this.}
The thermal energy 
with which jet material is injected is over $6$ orders of magnitude lower than the injected kinetic energy,
so completely negligible. 
The jets are also given an additional velocity component equal to the instantaneous orbital velocity of the companion.
For simplicity, the jet mass outflow rate and power are kept constant throughout a given simulation
and the jet is not made to rotate about its axis.
To conserve mass, the mass added to the jet is subtracted from the companion point particle.
Thus, in this work, as in \citetalias{Zou+22}, 
we do not attempt to model the dependence of the jet properties on other quantities such as the accretion rate;
such investigations are left for future work.

We emphasize that jet material is continuously added in the jet launch region (spherical sectors)
without removing the envelope gas that is already present there. 
Thus,
the density and velocity of gas in the jet launch region can take on any values 
and need not be similar to the density and velocity with which the jet is initialized in the subgrid model.
As the NS jets are initialized when the companion is already surrounded by dense gas (see Section~\ref{sec:runs}),
the jets are initially choked and their energy output thermalized. 
As more and more energy and momentum are added by jets, 
the jets can push their way outward by displacing overlying envelope material.
MS and WD jets are not powerful enough for such breakout to occur \citepalias{Zou+22},
which motivates us to explore the somewhat speculative case of highly super-Eddington NS jets, 
which are much more powerful.

\subsection{Tracking Jet, Star and Ambient Gas Components}
We use tracer particles to identify the material originating from the jets.
This allows us to plot the mass or energy density of the jet material separately from the other gas.
Likewise, we track the ambient medium so that it can be excluded in certain calculations.
We also separately track the remainder of the gas, hereafter referred to as the envelope gas.

\subsection{Companion Accretion}
For some runs, we employ a sub-grid accretion model for the secondary \citep{Krumholz+04}, 
which is independent of the subgrid jet model.
Within a radius of four cells of the companion, equal to $r\acc = 4\delta_4 = 0.56\Rsun$,
a certain fraction of mass determined from a condition based on the Bondi-Hoyle-Lyttleton formalism is removed from
the simulation domain and added to the point particle each time step \citep{Chamandy+18}. 
Note that $r\acc$ is chosen to be much smaller than $r\soft$ and $r\jet$ 
to ensure that the flow outside of the softening sphere is not directly affected by the accretion,
as well as to prevent most of the jet material from being accreted as soon as it is added.
The accretion rates in the simulations can be highly super-Eddington and we consider them to be upper limits.
Thus, to bracket the range of possibilities, we choose to perform runs with and without subgrid accretion turned on.

\subsection{Runs}\label{sec:runs}
The simulations were restarted from an earlier configuration of the WD run of \citetalias{Zou+22} (their Run~J8),
with $\dot{M}\jet$ and $v\jet$ modified suddenly upon restart to values appropriate for a NS companion.
The motivation for this choice was simply to make effective use of the computational resources available, 
since starting the simulation from $t=0$ with a high-speed jet is extremely computationally expensive,
and of the runs performed by \citetalias{Zou+22}, 
the parameter values of Run~J8 were closest to those of our NS runs.
We choose to turn the jets on suddenly rather than transitioning gradually between parameter values
for simplicity and ease of interpretation, but this aspect of the model could be refined in future work.
In any case, given that a WD jet has only a modest impact on the CE phase \citepalias{Zou+22}
and the accretion rate only becomes significant at $t\approx14\da$ \citep{Chamandy+18}, 
the assumption that the NS jet turns on at or a few days after $t=13.8\da$ seems appropriate.
We try restarting from Run~J8 at two different times, $t\restart=13.8\da$ and $19.9\da$, in the various runs
to explore how the results are affected by this parameter.

A summary of the runs performed is presented in Table~\ref{tab:runs}.
All runs were ended at $t=40\da$ to maximize the computational resources available
and because we have confidence that the numerical resolution
is sufficient up to this point but not necessarily beyond it \citep{Chamandy+19a}.
Runs~01, 03, 04, 05, 06, 07 and 09 were restarted from J8 from $t=19.9\da$, 
whereas Runs~08 and 11 were restarted from $t=13.8\da$. 
Run~01 is the same as J8 except that $\dot{M}\jet=2\times10^{-4}\Msunyr$, 
which is $10$ times higher than in J8, 
allowing us to isolate the effect of the jet mass-loss rate.
All other runs have $\dot{M}\jet=2\times10^{-4}\Msunyr$ and $v\jet\approx3\times 10^4\kms\approx0.1c$. 
Runs~03 and 05 are identical except that in Run~05 we reduce the density floor 
from $10^{-10}\gcmcmcm$ to $10^{-12}\gcmcmcm$ at $t=33.8\da$ to prevent the lowest densities from hitting the floor.
The effect on the results is negligible, so these two runs are identical from a practical point of view.
Runs~04 and 11 are identical to Runs~05 and 08, respectively, 
except that Runs~04 and 11 have subgrid accretion turned off.
This allows us to isolate the effects of accretion.
Run~09 is similar to Run~05 but the jet mass-loss rate was increased by an order of magnitude.
Finally, in Runs~06 (07) we reduced (increased) the maximum resolution by a factor of two from Run~05
in order to test the dependence of the simulation results on the numerical resolution.

\subsection{Unbound Mass Condition}\label{sec:unbound}
There is no single, standard definition of unbound mass because such a definition depends on whether 
and to what extent the thermal and self-gravitational potential energy of (bound and unbound) envelope gas are included,
and whether or not the potential energy nominally attributed to the particles is included 
\citep[e.g.,][]{Ivanova+13a,Chamandy+19a,Prust+Chang19,Chamandy+21err}.
Herein, we adopt the same criterion as in \citetalias{Zou+22}, 
which includes the thermal energy density, self-gravitational potential energy density considering all gas, 
and the full potential energy associated with the particle-gas interactions.
Qualitative results are not expected to depend on the precise choice of the condition for gas to be unbound
\citep{Chamandy+19a,Chamandy+24}.

\section{Results}\label{sec:results}

\subsection{Jet Morphological Evolution and Breakout}\label{sec:breakout}
Snapshots showing slices through the plane perpendicular to the orbital plane that contains both particles
(primary core and companion) are shown in Figs.~\ref{fig:run05}, \ref{fig:run05_zoom} and \ref{fig:run08_zoom}. 
Each row shows a different time and each column shows a different quantity:
envelope gas density (excluding jet and ambient densities), mass-weighted binding energy density,
sonic Mach number, and density of jet material.\footnote{Movies are available at 
\url{https://doi.org/10.5281/zenodo.21982012}.}
Figure~\ref{fig:run05} contains plots for Run~05 and shows the global structure of the envelope, 
Fig.~\ref{fig:run05_zoom} is similar to Fig.~\ref{fig:run05} but zooms in by a factor of three, 
and Fig.~\ref{fig:run08_zoom} is similar to Fig.~\ref{fig:run05_zoom} but now for Run~08 (early-onset NS jets).
The two circles near the center of each panel 
show the gravitational softening spheres of the RGB core (left) and companion (right).

From the left column of Fig.~\ref{fig:run05}, 
we see that the density of envelope material in a bi-conical region below and above the companion steadily decreases,
owing to the propagation of the low-density jets (right column).
However, by the end of the simulation these regions are still not completely evacuated by the jets.
From the middle columns, 
we see that the envelope material in the bi-conical region is unbound and supersonic,
though the magnitude of its binding energy compared to adjacent bound material in the envelope is relatively low
due to its low density.

In Fig.~\ref{fig:run05_zoom}, which is a zoomed-in version of Fig.~\ref{fig:run05}, 
we see that the jets are intermittent at first, but becomes steadier by the end of the simulation,
and appear to be in the process of breaking out of the envelope.
This is more apparent in the movies (which show Run~03 instead of Run~05).
However, it is possible that they would become more intermittent again at later times.

Thus, contrary to the MS and WD jets studied in \citetalias{Zou+22},
the NS jets strongly affect the envelope morphology as they are able to drill through the envelope
and partially break out during the timescale of the simulation.

Figure~\ref{fig:run08_zoom} shows Run~08, 
which is otherwise identical to Run~05 except that the jets are turned on at $t=13.8\da$ rather than $t=19.9\da$.
The jet properties are similar to those in Run~05 but the jets remain intermittent up until the end of the simulation.

In both runs, the jets can sometimes be highly asymmetric with respect to one another, as seen, for example,
in the bottom row of Fig.~\ref{fig:run05} or the top row of Fig.~\ref{fig:run08_zoom}.
This shows that small perturbations and randomness in the initial conditions can lead to large asymmetries in jet structure.

\begin{figure}
  \centering
   \includegraphics[width=\columnwidth]{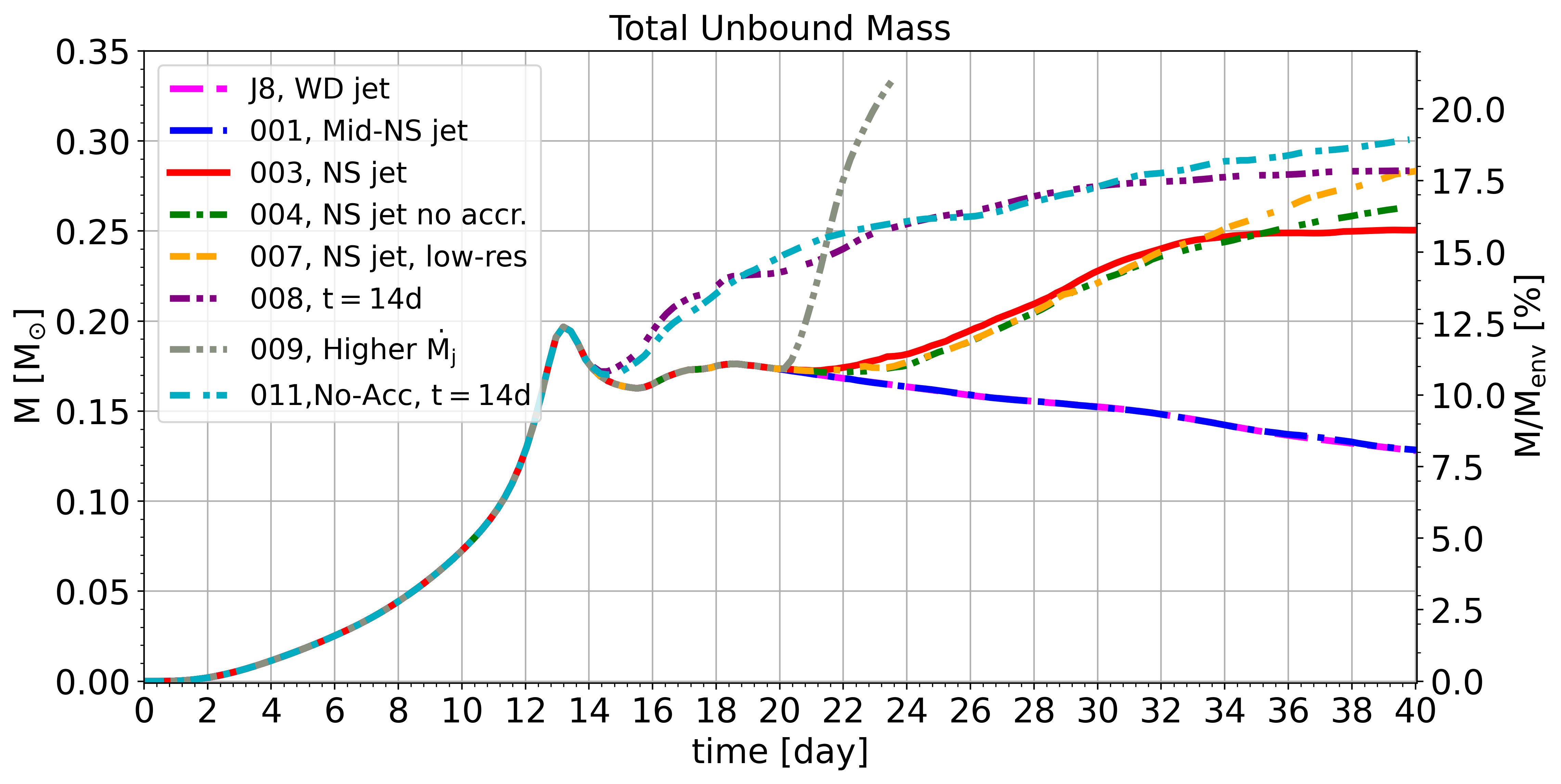}\\
   \includegraphics[width=\columnwidth]{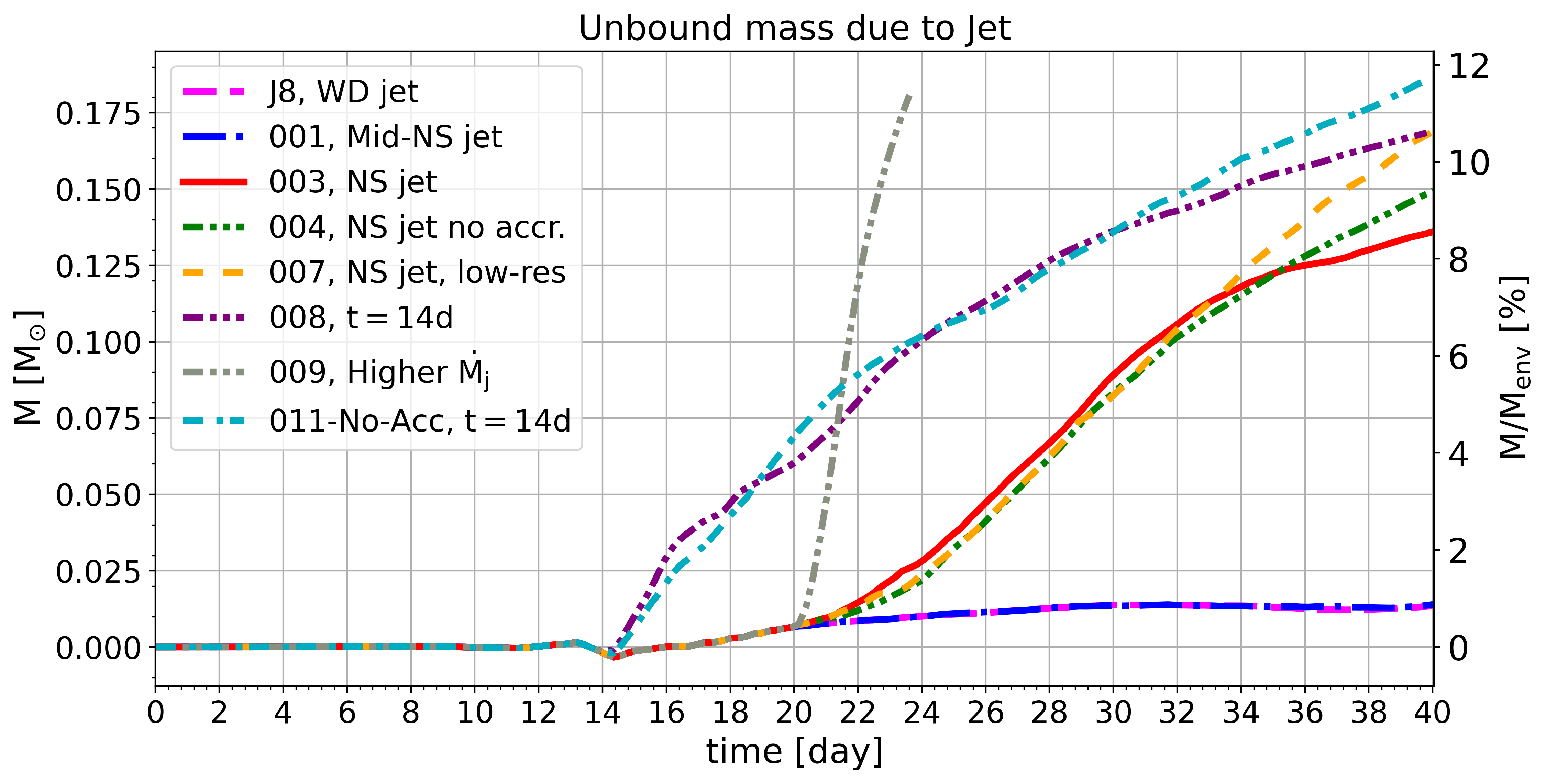}
\caption{Top: Evolution of the mass of the unbound envelope gas, relative to the initial value.
  Bottom: Evolution of the difference between the mass of the unbound envelope gas in a given run 
  and that in the run without a jet (Run~NJ1). This difference is essentially equal to the mass unbound by the jet. 
  Note that some of the model names in the legend prepend the model names in Table~\ref{tab:runs} by a zero.}
\label{fig:unbound}
\end{figure}

\subsection{Envelope Unbinding}\label{sec:unbinding}
In the top panel of Fig.~\ref{fig:unbound}, 
we show the total mass of unbound envelope gas as a function of time.
In the bottom panel, 
the unbound envelope mass from Run~NJ1 has been subtracted off, 
leaving the amount of unbound envelope mass that was unbound by the jets.%
\footnote{This statement assumes that subgrid accretion does not directly affect the amount of unbound mass significantly,
which is reasonable since the mass that is accreted tends to be strongly bound.}

In the WD run J8, the jets unbind about $1\%$ of the total envelope mass, 
which is an extra $\sim10\%$ as compared to the mass unbound by the transfer of orbital energy alone, 
as measured from Run~NJ1, which does not have jets \citepalias{Zou+22}.
From Fig.~\ref{fig:unbound}, 
we see that the unbound mass in Run~01 (blue), 
where $v\jet$ is the same as in Run~J8 but $\dot{M}\jet$ is larger by an order of magnitude,
is about the same as that in Run~J8 (magenta).
In J8, the jet is barely visible in the snapshots similar to those in Figs.~\ref{fig:run05}--\ref{fig:run08_zoom}
and associated movies, but in Run~01 it is clearly visible in the movies.
Thus, despite the more powerful jets of Run~01 having a noticeable influence on the envelope morphology,
their effect on the unbound mass is modest.

All of the other runs unbind considerably more mass, 
and the mass unbound by the jets, of order $10\%$ of the envelope mass (Fig.~\ref{fig:unbound}, bottom panel), 
is comparable to the mass unbound by orbital energy.
However, the rate of unbinding (slope) decreases with time.
Apparently, as the jets break out from the envelope, 
they interact less and less with it, and hence are less and less efficient at unbinding material.
This suggests that there may be an asymptotic limit to the unbinding rate by jets.
As might be expected, the runs for which the NS jet is turned on earlier unbind more material overall.
However, the difference in the unbound mass between Runs~08 and 03, 
or between Runs~11 and 04, decreases with time. 
This is understandable if jets that turn on earlier also break out earlier.
Roughly speaking, the unbound mass curve is effectively translated in time by the difference in restart times ($6\da$).

Run~09 (10~times larger jet power) unbinds material at a much higher rate, 
as seen in the bottom panel of Fig.~\ref{fig:unbound}. 
The initial shape of the curve, including the decreasing rate, is similar to that seen for the other runs.
Comparing this curve with the part of the curve for Run~03 that is qualitatively similar 
(up to a point not far beyond the inflection point),
we estimate a roughly $6$-fold increase in the mass unbinding rate.
Run~09 could not be extended further in time due to a lack of computational resources.

As the jets seem to be breaking out of the envelope but have still not fully broken out,
we would expect the mass unbinding rate to decrease further at late times after the simulations end.
In Section~\ref{sec:extrapolation}, 
we explore the role played by the jets in envelope unbinding over the course of the entire CE phase.

\subsubsection{Role of Accretion}
Comparing Runs~05 with Run~04, for which subgrid accretion was turned off,
we see that the unbound mass at the end of the simulation is slightly larger for the no accretion case.
Runs~08 and 11, which restarted from J8 at $t=13.8\da$, show a similar difference, 
with the no accretion case leading to slightly more unbinding.
Accretion helps to evacuate the conical regions, facilitating jet propagation through the envelope.
When accretion is absent, the jets cannot break out as easily, 
prolonging their interaction with the envelope and increasing envelope unbinding.

\subsubsection{Effects of Changing the Resolution}
For Run~07, which was performed at lower maximum resolution, 
we see the same rate of unbinding at early times but a larger rate at late times.
Comparison of Runs~07 and 05 suggests that any improvement in the resolution would cause the unbound mass 
curve to flatten even more.
Thus, the conclusion that the unbinding rate is quenched as the jet breaks out seems to be robust 
to changes in the numerical resolution.
The morphology in the different-resolution runs is compared in Appendix~\ref{sec:resolution}.

\begin{figure}
    \centering
    \includegraphics[width=1.0\columnwidth, clip = true,trim={0 10 0 0}]{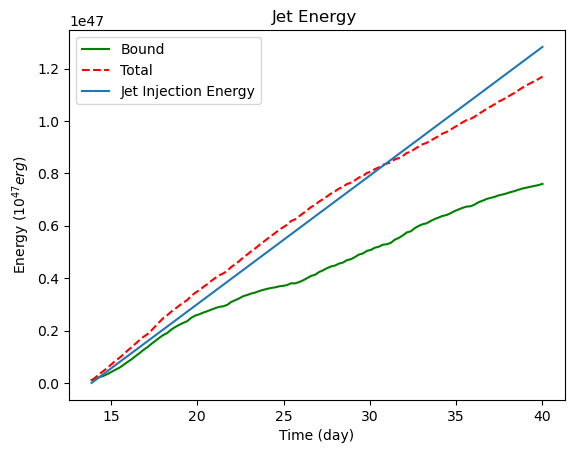}
    \caption{Energy transfer by jets in the simulation. 
      The solid blue line shows the approximate energy $\tfrac{1}{2}\dot{M}\jet v\jet^2(t-t\restart)$ 
      injected by the jets in Run~11 (early-onset NS jets without accretion). 
      The dashed red line shows the difference in the energy in the envelope gas (both bound and unbound)
      between Runs~11 and NJ1. 
      The solid green curve shows the change in the difference in energy contained in bound gas between Run~11 
      and Run~NJ1 (no jet, no accretion).
      The deficit of the green line with respect to the blue line 
      is mainly caused by energy transfer from the jet to already unbound gas,
      which reduces the efficiency of envelope unbinding.
    \label{fig:jet_energy}
    }
\end{figure}

\subsection{Energy Transfer from Jets to Envelope Gas}
\begin{figure*}
  \centering
  \includegraphics[width=0.8\linewidth]{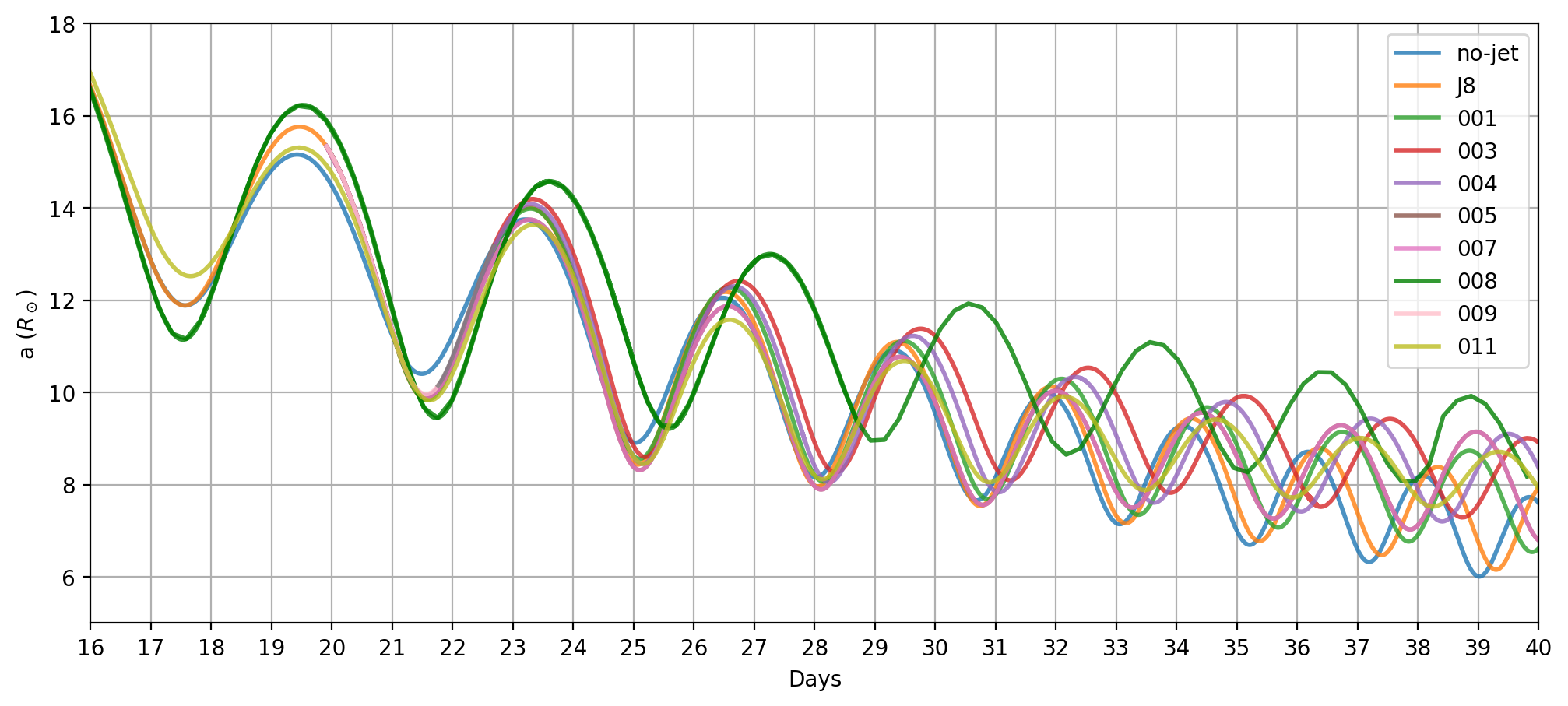}
\caption{Evolution of the orbital separation between the RGB core and companion point particles,
  for the runs performed. The label ``no-jet'' refers to Run~NJ1.}
\label{fig:separation}
\end{figure*}

In Fig.~\ref{fig:jet_energy}, 
the solid blue line shows the approximate energy injected by the jet $\tfrac{1}{2}\dot{M}\jet v\jet^2(t-t\restart)$ 
as a function of time,
for Run~11 (early-onset NS jet without accretion). 
The dashed red line shows the difference in energy in the envelope gas 
(both unbound and bound) between Run~11 and Run~NJ1 (no accretion, no jets);
the decrease in slope at $t\approx28\da$ is caused by 
the jet retaining its energy or transferring energy to the ambient medium instead of the envelope.
The green line shows the difference in the energy contained in the \textit{bound} gas between the two runs.
Recall that bound gas is defined as that which has negative energy density at time $t$ (Section~\ref{sec:unbound}).
Transferring jet energy to bound gas helps to unbind the envelope, while transferring it to unbound gas does not.
In the first five days following jet activation, 
essentially all of the jet energy is transferred to bound gas.
This is reasonable because the jet is still confined to the jet activation region, where the gas is tightly bound.
However, the rate of energy transfer to bound gas decreases significantly at $t\approx20\da$, 
which suggests that jet energy is being transferred less efficiently
because a significant fraction is being transferred to unbound gas.%
\footnote{After $t=32\da$, 
the orbital separation evolution curve of Run~NJ1 is slightly steeper than that of Run~11 (Fig.~\ref{fig:separation}),
which likely results in a higher rate of orbital energy transfer to bound gas in Run~NJ1 than in Run~11.
Thus, the difference in the energy contained in the bound gas between Runs~11 and NJ1 cannot 
wholly be attributed to energy deposited by the jet, 
and the difference would in fact be slightly larger if not for this extra orbital energy injection in Run~NJ1.
For this reason, the green curve in Fig.~\ref{fig:jet_energy}
may slightly underestimate the rate of jet energy transfer into bound gas in Run~11 after $t=32\da$.} 

To summarize our simulation results pertaining to envelope unbinding,
we find that the NS jets unbind envelope mass at a rate comparable to that due to the transfer of orbital energy alone,
and at an even larger rate for the model with the most powerful jets (Run~09). 
However, there is a negative feedback effect present: 
as the jets drill through the envelope on their way to breaking out, 
their interaction with the envelope weakens,
resulting in a decrease in the rate of mass unbinding.

\begin{figure}
 \centering
 \includegraphics[width=\columnwidth]{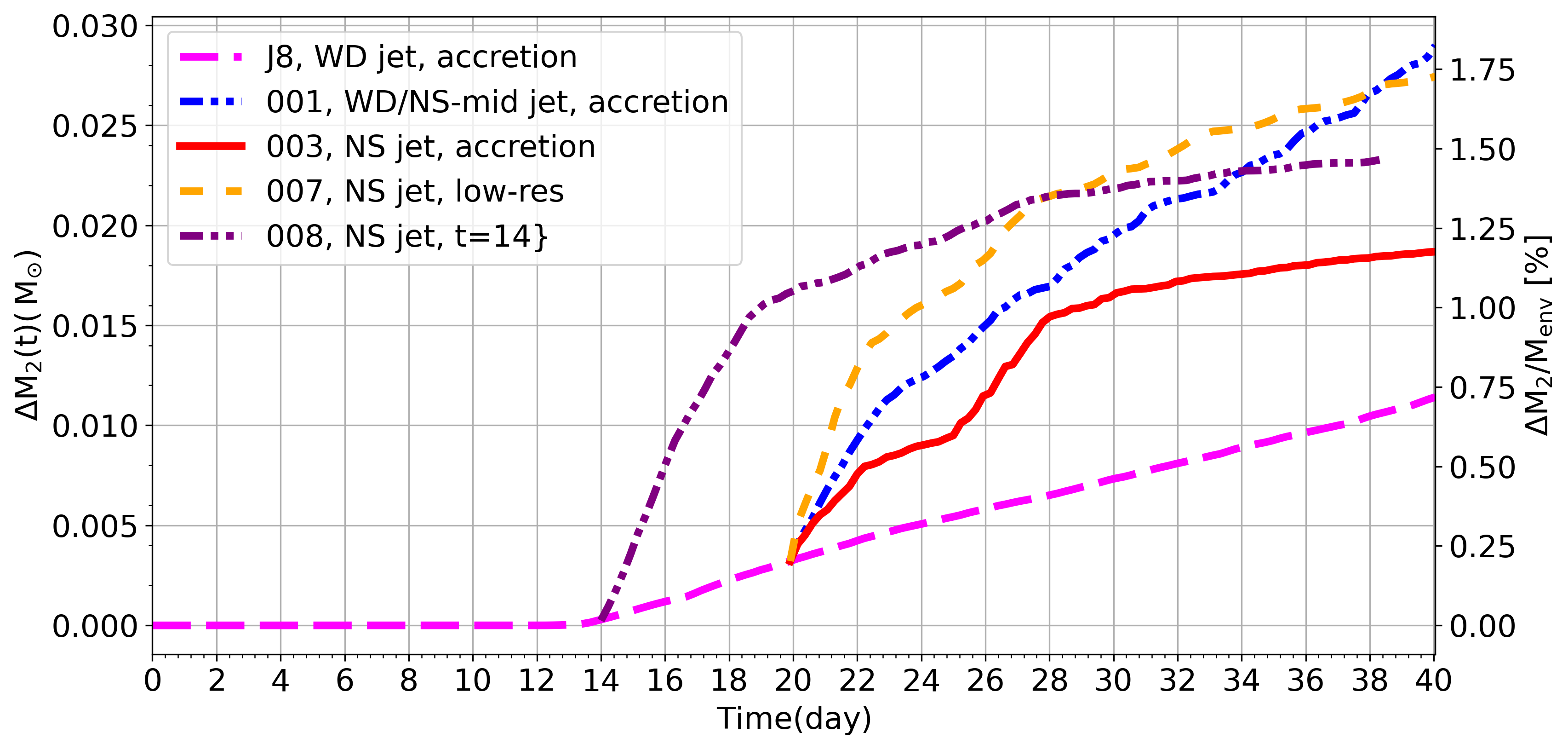}
\caption{Evolution of the mass accreted by the point-particle companion for a subset of runs.
\label{fig:accretion}
}
\end{figure}

\subsection{Orbital Evolution}\label{sec:orbit}
Run~05 (late-onset NS jets) starts showing differences in envelope morphology from Run~J8 (WD) 
almost immediately after the NS jet is turned on.
The jets start breaking out around day $33$ and by day $36$ have largely broken out 
and remain so until the end of the simulation at $t=40\da$. 
Aside from reducing the efficiency of envelope unbinding, 
jet breakout also affects the orbital motion of the particles, causing the tightening of their separation to slow. 
The orbital separation for all of the runs is presented in Fig.~\ref{fig:separation}.
We see that after $t\approx32\da$, 
all the runs have slightly shallower separation curves compared to the no-jet, no-accretion run NJ1 (blue).
This could be because the jets remove material from the vicinity of the secondary, 
resulting in a decrease in the gravitational drag, 
which in turn leads to a decrease in the rate of orbital energy transfer. 
Thus, there is a a second type of negative feedback that may affect envelope unbinding:
the transfer of jet energy to the envelope causes a slight reduction in the rate of orbital energy transfer to the envelope.

\subsection{Accretion Rate}\label{sec:accretion_rate}
For simulations for which subgrid accretion is turned on, 
mass is accreted onto the companion at super-Eddington rates, adding to its mass.
This results in the removal of more than $1\%$ of the envelope mass, as shown in Fig.~\ref{fig:accretion}. 
In Run~J8 (WD), the accretion rate reaches the imposed upper limit, 
but no such ceiling is imposed in the NS jet runs that include accretion.
We see that the accretion rate decreases markedly with time in these runs.
This is caused by feedback: the jet removes material from the vicinity of the secondary, 
lowering the density there and resulting in a lower accretion rate.


\section{Beyond the Present Model}
\subsection{Jet Model}\label{sec:caveats}
Apart from the question of whether highly super-Eddington jets can be supported during a CE event, 
the model has a few caveats that could be addressed in future work.
First, the jet mass-loss rate and power are kept constant, 
whereas in reality, they would be affected by certain parameters including the accretion rate.
It would thus be interesting to make the jet properties depend on the properties of the accreting matter.
For instance, the mass-loss rate $\dot{M}\jet$ could be set to be a fixed fraction of the accretion rate.
This could result in yet another negative feedback mechanism 
if the reduction in the accretion rate owing to jet breakout (Fig.~\ref{fig:accretion})
leads to weaker jets, which in turn allows the accretion rate to recover, 
leading to cyclical behaviour \citep{Soker16,Morenomendez+17,Lopez-camara+19,Lopez-camara+20}.
We have seen that in the present model the jets can sometimes be intermittent~--
they might be more intermittent if this feedback were included.
However, it is interesting that such feedback does not seem to be a necessary condition for intermittency.

Furthermore, 
some small fraction of the accreted angular momentum could be transferred to the jets
and along with stochastic jet asymmetries across the midplane \citep{Pettibone+2009}
could together cause the jets to spin or wobble about their axis and the orientation of this axis to vary with time.
If the amplitude of this angular variation were sufficiently high,
then this effect might help to keep the jets and envelope energetically coupled, 
preventing or delaying breakout and increasing the efficiency of envelope unbinding 
(by so-called wobbling, or jittering jets; \citealt{Papish+Soker11}).

\subsection{Extrapolating the Unbound Mass Evolution}\label{sec:extrapolation}
For practical reasons, the simulations are terminated relatively early during the CE phase,
but it is important to understand the long-term effects of jets on envelope unbinding. 
Thus, we now turn to order-of-magnitude estimates in order to crudely extrapolate our results to late times.

\citet{Chamandy+24} performed four simulations with essentially the same initial conditions as in this work,
but excluding accretion and jets, 
and with higher resolution and longer durations of between $50\da$ and $142\da$.
Two of the simulations adopted an ideal gas equation of state and the other two adopted tabulated equations of state.
For the three simulations that extended beyond $t=70\da$, 
the mass unbinding rate was found to be roughly constant after that time, 
and almost independent of the equation of state.
By way of simple extrapolation, assuming that the mass unbinding rate would remain constant,
\citet{Chamandy+24} roughly estimated that the transfer of orbital energy alone 
could unbind the envelope in about $2\yr$ for this system.
Does the presence of NS jets leads to a different estimate?

In Run~03, the jets begin to break out around day~$35$, 
and thereafter unbind the envelope at a rate of about $0.2\%$ per day (top panel of Fig.~\ref{fig:unbound}).
At this rate, the remaining roughly $85\%$ of the envelope will be unbound in a little over $1\yr$.
However, this is a lower limit because the unbinding rate is expected to decrease with time as the jets break out
and continue to decouple from the envelope gas.
Even so, this tells us that the very powerful jets considered in this work might contribute to envelope 
unbinding at the same level as the transfer of orbital energy.
However, the reduction in the drag (Section~\ref{sec:orbit}), and thus in the rate of orbital energy transfer, 
caused by the jets is a negative feedback effect that counteracts (though perhaps only mildly) the unbinding from jets.
In any case, this is a best-case scenario for the unbinding ability of the highly super-Eddington NS jets considered
(Table~\ref{tab:runs}), perhaps with the exception of Run~09, 
which assumes a jet mass-loss rate and jet power one order of magnitude larger still.

For choked jets, the jet energy is thermalized inside the envelope, 
allowing the jet energy to be utilized to unbind envelope material.
Once the jets break out, 
this no longer occurs and the jet energy largely escapes.
However, energy may still be transfered from the jets
to the envelope at a lower rate, e.g., via turbulent diffusion through a thin boundary later.
We now provide an order-of magnitude estimate of the effect this may have on envelope unbinding.
It is possible to estimate the rate of energy transfer and the timescale to unbind the envelope,
but such estimates are uncertain because they are sensitive to the magnitude of the velocity fluctuations, 
which varies strongly across the boundary layer.
Nevertheless, 
we can model the energy transfer rate due to turbulent transport as
\begin{equation}
 \dot{E}\turb \approx \frac{\rho V v\turb^3}{2l},
\end{equation}
where $v\turb$ is the rms turbulent speed in the envelope adjacent to the boundary layer,
$\rho$ is the gas density,
$V \approx H[\uppi R^2 - \uppi(R-l)^2]$ is the volume of the roughly cylindrical boundary layer,
$R$ its radius, $H$ its height, and $l$ its thickness.
Dividing the approximate magnitude of the velocity fluctuations in or near the boundary layer by the local sound speed $c\sound$, 
we estimate the turbulence to be mildly subsonic, and use the local sound speed as a proxy for $v\turb$.
We estimated these parameters for a slice through the particles near the end of Run~03, obtaining
$v\turb\approx10$--$30\kms$, $\rho\approx10^{-5}\gcmcmcm$, $R\approx2\Rsun$, $l\approx1\Rsun$ and $H\approx25\Rsun$.
At this time, the remaining envelope binding energy is $E\approx 10^{47}\erg$ \citep{Chamandy+19a},
so the unbinding time can be estimated as $E\bind/\dot{E}\turb \approx 20$--$500\yr$.

While very rough, 
the estimates discussed in this subsection suggest that the role played by such jets in unbinding the envelope after jet breakout
may be subdominant to that played by orbital energy transfer.
By contrast, 
if the jet energy could somehow be used with perfect efficiency to unbind the envelope, 
then the envelope would be unbound in $E\bind/\dot{E}\jet \approx 0.1\yr$.
The disparity in these estimates illustrates that one should be cautious when making simple analytic estimates
about the role of jets in CE evolution, particularly strong jets that break out.

\section{Conclusions}\label{sec:conclusions}
In this work we have explored the effects of jets launched by NS companions orbiting within common envelopes
involving a $2\Msun$ RGB star, with a focus on assessing the role played by such jets in unbinding the envelope.
Our 3D global hydrodynamical simulations include self-gravity, model the RGB core and companion as point particles,
and are terminated after $t=40\da$ to maximize computational resources.
Jets are launched from the companion using a subgrid model that conserves mass, momentum, and angular momentum. 
Accretion onto the companion is controlled independently by a separate subgrid model, 
and was turned on for a subset of runs; 
its direct effect on envelope unbinding is found to be subdominant compared to that of the jets.
The jets are assumed to have constant mass outflow rates $\dot{M}\jet$ 
that exceed the Eddington rate by roughly four orders of magnitude,
and powers (dominated by bulk kinetic energy) of $\tfrac{1}{2}\dot{M}\jet v\jet^2$, with $v\jet = 0.1c$.
Such an extreme scenario might occur if NS accretion is mediated by neutrinos,
or if most of the accreting mass is channeled directly to jets.
The key results of the study are as follows:
\begin{enumerate}
  \item Unlike jets launched from a main sequence star or WD, 
    which are choked inside the envelope \citepalias{Zou+22},
    the NS jets simulated pierce their way through the envelope and in fact begin to completely break out by the ends of the simulations.
  \item This results in a bipolar CE morphology with hot, low-density lobes moving outward at speeds 
    of hundreds to thousands of $\!\kms$.
  \item Despite the jet input power and mass-loss rate being constant in our model, 
    the outward propagation of jet material through the envelope can sometimes be intermittent, 
    resulting in disconnected and asymmetric jets (Section~\ref{sec:breakout}).
    However, at other times, the jets are continuous and symmetric, 
    though they tend to curve away from the jet axis (Figs.~\ref{fig:run05}-\ref{fig:run08_zoom}).
  \item During the simulation the jets are responsible for unbinding 
    about the same amount of envelope mass as the orbital tightening of the core and NS particles (Fig.~\ref{fig:unbound}).
  \item However, 
    the jets unbind the envelope at lower and lower rates as they propagate to larger distances and eventually break out~--~
    this self-regulating behaviour is a consequence of the reducing rate of energy transfer as the jets break out.
  \item We also find a secondary, more subtle negative feedback effect: 
    jet activity leads to a reduction in the drag and thus a slightly lower rate of orbital energy transfer
    compared to the simulations without jets, which tends to reduce the amount of unbinding caused by orbital tightening
    (Section~\ref{sec:orbit}).
  \item Beyond the $40\da$ timescale of the simulations, 
    the jets would continue to transfer energy to the envelope at a lower rate.
    For the jets considered, 
    we crudely estimate that this energy transfer is likely to be subdominant as compared to transfer from orbital energy,
    but longer simulations are needed to study the ultimate fate of such systems.
\end{enumerate}
This work shows that very high-powered NS jets have a lesser role in unbinding the envelope than 
simple estimates that neglect jet breakout would suggest.
However, our simulations have fairly short durations and neglect effects like jet precession or jittering \citep{Papish+Soker11}, 
which could enhance the coupling between the jets and envelope, leading to more efficient unbinding.
Angular variability of the jet axis could arise naturally by allowing jets to acquire angular momentum from accreted material.
We have also kept the jet power constant, even though it should depend on the accretion rate,
which is itself affected by the jets, 
so a more realistic treatment would incorporate such a feedback loop;
this has been attempted for wind tunnel-type CE simulations \citep{Lopez-camara+19}.
Also, jet-launching CE events involving high-mass evolved stars might lead to different results, 
and should be studied in the future.
Current constraints do not preclude the existence of even stronger jets from super-Eddington accretion
of CE gas onto NS or black hole companions, 
or, perhaps more plausibly, from a merger between the core and companion or between the components of a companion binary
\citep[e.g.,][]{Soker21,Grichener+Soker23}.

Overall, this work and that of \citetalias{Zou+22} together show that while jets can, in principle, 
affect the CE phase, those with powers significant enough to temporarily dominate envelope unbinding
would quickly break out of the envelope, limiting their own influence,
although the long-term consequences of this feedback remain somewhat uncertain.

\section*{Acknowledgments}
The authors are grateful to the reviewer for a constructive and thorough review, 
and to Jonathan Carroll-Nellenback, Nishikanta Khandai, and Tuhin Ghosh for discussions.
We acknowledge the use of the Astro cluster at the National Institute of Science Education and Research (NISER), India,  
supported by the Department of Atomic Energy grant RIN-4001,
and the Cori supercomputer at the National Energy Research Scientific Computing Center (NERSC), 
a DOE Office of Science User Facility supported 
by the Office of Science of the U.S.~Department of Energy under Contract No.~DE-AC02-05CH11231,
through NERSC award FES-ERCAP m4008.
We also acknowledge support from NSF grant PHY-2020249 which funds the NSF Physics  Center for Matter at Atomic Pressure (CMAP).
JN acknowledges support from U.S.~National Science Foundation awards AST-2009713, AST-2319326 and AST-2511139.

\section*{Data and Code Availability}
The code and simulation output will be made available upon reasonable request to the corresponding author.

\appendix
\section{Resolution study}\label{sec:resolution}
Runs~06 and 07 were identical to Run~05 except for the maximum adaptive mesh refinement (AMR) resolution:
Run~05 uses the standard maximum resolution of $\delta_4=0.14\Rsun$ (maximum AMR level 4),
Run~06 uses a twice higher maximum resolution of $\delta_5=0.07\Rsun$ (maximum AMR level 5), 
and Run~07 uses a twice lower maximum resolution of $\delta_3=0.28\Rsun$ (maximum AMR level 3).
Here, $\delta_n$ denotes the cell size at AMR refinement level $n$.
Run~06 was run for a much shorter duration than other runs owing to limitations in the amount of computational resources. 

\subsection{Jets}
\begin{figure*}    
    \includegraphics[width=\textwidth]{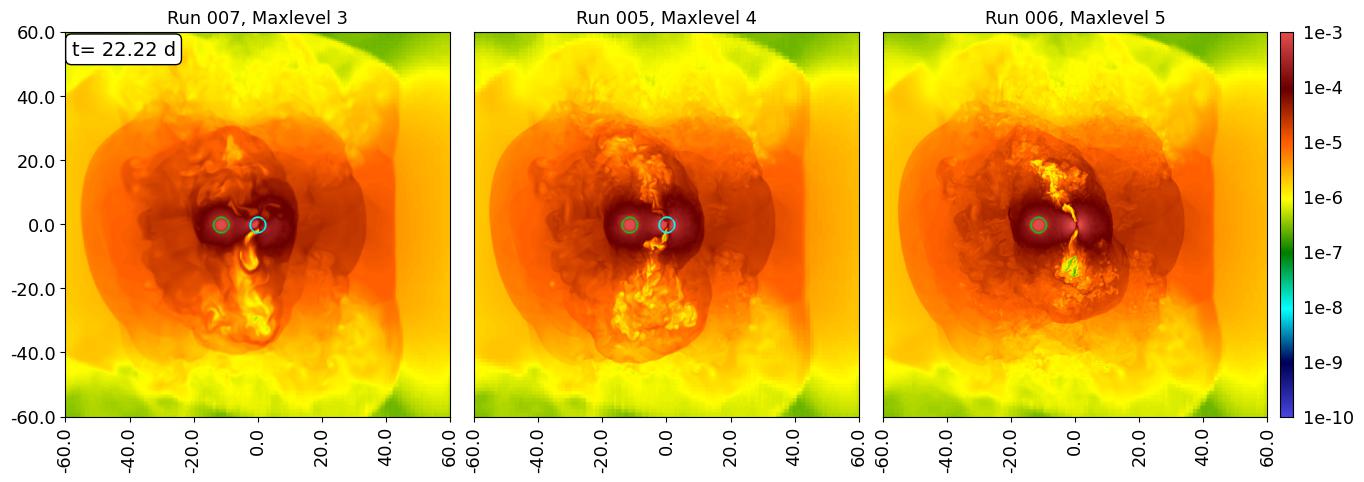}
    \includegraphics[width=\textwidth]{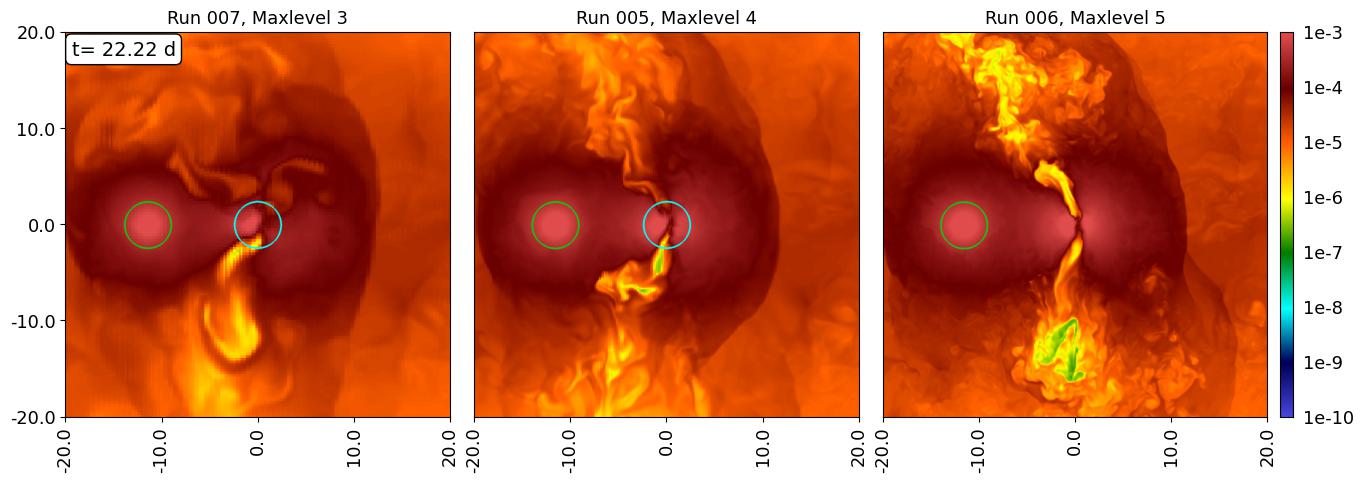}\\
    \caption{Comparison of a slice through the particles of the gas density for runs that are identical 
      except for the maximum resolution, 
      which improves by a factor of two between successive columns (from left to right).
      Rows show two different zoom levels, with axis units in $\!\Rsun$.
      While the envelope structure is very similar between the runs and the jet morphology is qualitatively similar,
      the detailed structure of the jets is different.
      This is not only due to differences in resolution per se, 
      but also to intermittency and stochasticity associated with the jets.
    \label{fig:jet_resolution}
    }
\end{figure*}

Figure~\ref{fig:jet_resolution} compares the gas density of the three runs at $t=21.76\da$,
about $2\da$ after the NS jets are initiated, at two different zoom levels.
While the detailed jet structure is markedly different between the runs, 
the overall morphologies of the envelope gas and jets are quite consistent.
The differences in jet structure are caused by the stochastic nature of the jets, 
whose detailed evolution is sensitive to small fluctuations in the initial conditions.
It can be seen, however, that the jets seem to break out more easily as the resolution improves,
presumably leading to less energy transfer from the jets to the envelope gas.
This is consistent with Fig.~\ref{fig:unbound}, 
which shows that the low resolution run (Run~07) unbinds envelope mass at a higher rate than Run~05 at late times.
Accordingly, the main conclusion of this work~-- 
that the role played by the jets in envelope unbinding is rendered inefficient by their breakout~--
seems to be strengthened by increasing the resolution.

\subsection{Orbital Evolution}
\begin{figure}
    \centering
    \includegraphics[width=\columnwidth]{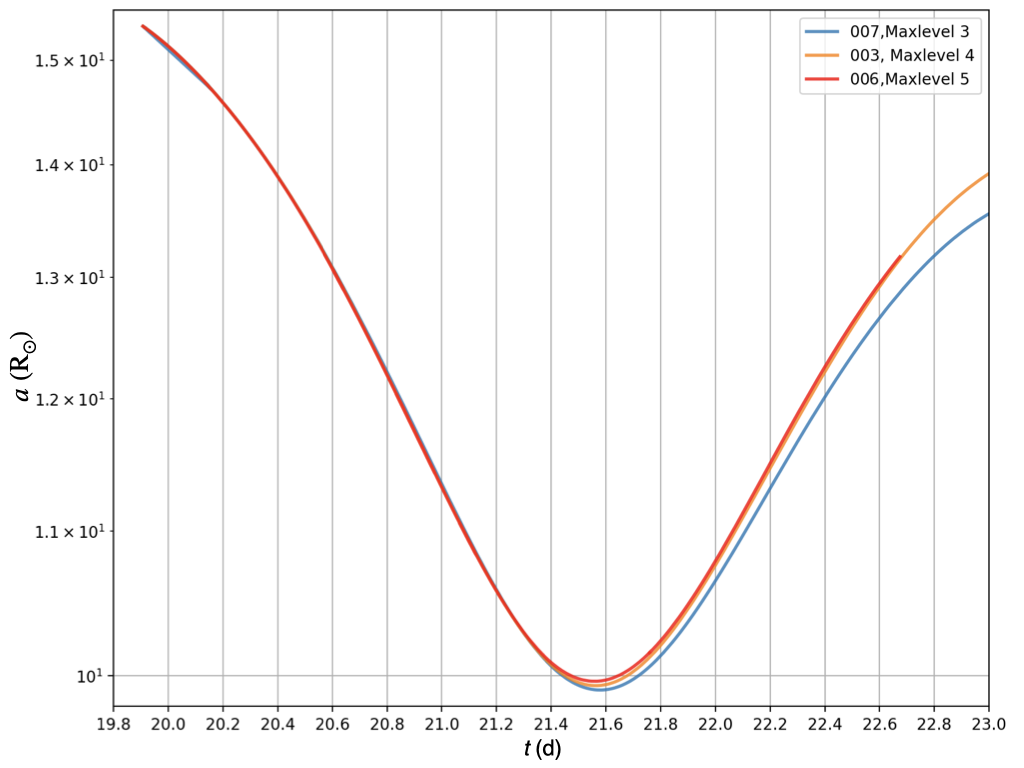}
    \caption{Resolution study comparing the evolution of the inter-particle separation for three runs: 
      Run~07 with maximum AMR level~3 in blue, 
      Run~05 with maximum AMR level~4 in orange (fiducial resolution), 
      and Run~06 with maximum AMR level~5 in red.
      The proximity of the orange and red lines suggests that the orbital evolution is well converged with resolution
      for the majority of runs, which utilize 4 AMR levels.
      }
    \label{fig:sep_resolution}
\end{figure}

Figure~\ref{fig:sep_resolution} shows the evolution of the inter-particle separation for Runs~07, 05, and 06,
which are identical except that they utilize 3, 4, or 5 levels of AMR, respectively.
During the $3\da$ duration of Run~06,
we see that the separation curve for all three runs is similar but that Run~07 shows a relatively large deviation
as compared to Runs~05 and 06, which have separation curves that are very close together.
These results demonstrate that the orbital evolution is reasonably well converged with resolution 
for our standard resolution of $4$ AMR levels.

\bibliography{refs}{}

\begin{thebibliography}{}
\expandafter\ifx\csname natexlab\endcsname\relax\def\natexlab#1{#1}\fi
\providecommand{\url}[1]{\href{#1}{#1}}
\providecommand{\dodoi}[1]{doi:~\href{http://doi.org/#1}{\nolinkurl{#1}}}
\providecommand{\doeprint}[1]{\href{http://ascl.net/#1}{\nolinkurl{http://ascl.net/#1}}}
\providecommand{\doarXiv}[1]{\href{https://arxiv.org/abs/#1}{\nolinkurl{https://arxiv.org/abs/#1}}}

\bibitem[{P.~J. {Armitage} \& M. {Livio}(2000){Armitage} \&
  {Livio}}]{Armitage+Livio00}
{Armitage}, P.~J., \& {Livio}, M. 2000, \bibinfo{title}{{Black Hole Formation
  via Hypercritical Accretion during Common-Envelope Evolution},} \apj, 532,
  540, \dodoi{10.1086/308548}

\bibitem[{J.~J. {Carroll-Nellenback} {et~al.}(2013){Carroll-Nellenback},
  {Shroyer}, {Frank}, \& {Ding}}]{Carroll-Nellenback+13}
{Carroll-Nellenback}, J.~J., {Shroyer}, B., {Frank}, A., \& {Ding}, C. 2013,
  \bibinfo{title}{{Efficient parallelization for AMR MHD multiphysics
  calculations; implementation in AstroBEAR},} Journal of Computational
  Physics, 236, 461, \dodoi{10.1016/j.jcp.2012.10.004}

\bibitem[{L. {Chamandy} {et~al.}(2024){Chamandy}, {Carroll-Nellenback},
  {Blackman}, {Frank}, {Tu}, {Liu}, {Zou}, \& {Nordhaus}}]{Chamandy+24}
{Chamandy}, L., {Carroll-Nellenback}, J., {Blackman}, E.~G., {et~al.} 2024,
  \bibinfo{title}{{How negative feedback and the ambient environment limit the
  influence of recombination in common envelope evolution},} \mnras, 528, 234,
  \dodoi{10.1093/mnras/stae036}

\bibitem[{L. {Chamandy} {et~al.}(2019){Chamandy}, {Tu}, {Blackman},
  {Carroll-Nellenback}, {Frank}, {Liu}, \& {Nordhaus}}]{Chamandy+19a}
{Chamandy}, L., {Tu}, Y., {Blackman}, E.~G., {et~al.} 2019,
  \bibinfo{title}{{Energy budget and core-envelope motion in common envelope
  evolution},} \mnras, 486, 1070, \dodoi{10.1093/mnras/stz887}

\bibitem[{L. {Chamandy} {et~al.}(2021){Chamandy}, {Tu}, {Blackman},
  {Carroll-Nellenback}, {Frank}, {Liu}, \& {Nordhaus}}]{Chamandy+21err}
{Chamandy}, L., {Tu}, Y., {Blackman}, E.~G., {et~al.} 2021,
  \bibinfo{title}{{Erratum: Energy budget and core-envelope motion in common
  envelope evolution},} \mnras, 501, 2209, \dodoi{10.1093/mnras/staa3846}

\bibitem[{L. {Chamandy} {et~al.}(2018){Chamandy}, {Frank}, {Blackman},
  {Carroll-Nellenback}, {Liu}, {Tu}, {Nordhaus}, {Chen}, \&
  {Peng}}]{Chamandy+18}
{Chamandy}, L., {Frank}, A., {Blackman}, E.~G., {et~al.} 2018,
  \bibinfo{title}{{Accretion in common envelope evolution},} \mnras, 480, 1898,
  \dodoi{10.1093/mnras/sty1950}

\bibitem[{P.~C.-K. {Cheong} {et~al.}(2026){Cheong}, {Radice}, \&
  {Fryer}}]{Cheong+26}
{Cheong}, P. C.-K., {Radice}, D., \& {Fryer}, C.~L. 2026,
  \bibinfo{title}{{Hyperaccreting neutron stars inside massive envelopes: The
  implausibility of Thorne-{\.Z}ytkow objects},} \prd, 114, 023018,
  \dodoi{10.1103/8pb2-9t2x}

\bibitem[{R.~A. {Chevalier}(1993){Chevalier}}]{Chevalier93}
{Chevalier}, R.~A. 1993, \bibinfo{title}{{Neutron Star Accretion in a Stellar
  Envelope},} \apjl, 411, L33, \dodoi{10.1086/186905}

\bibitem[{C. {Federrath} {et~al.}(2014){Federrath}, {Schr{\"o}n}, {Banerjee},
  \& {Klessen}}]{Federrath+14}
{Federrath}, C., {Schr{\"o}n}, M., {Banerjee}, R., \& {Klessen}, R.~S. 2014,
  \bibinfo{title}{{Modeling Jet and Outflow Feedback during Star Cluster
  Formation},} \apj, 790, 128, \dodoi{10.1088/0004-637X/790/2/128}

\bibitem[{J. {Frank} {et~al.}(2002){Frank}, {King}, \& {Raine}}]{Frank+02}
{Frank}, J., {King}, A., \& {Raine}, D.~J. 2002, {Accretion Power in
  Astrophysics: Third Edition}, 398

\bibitem[{A. {Grichener} \& N. {Soker}(2023){Grichener} \&
  {Soker}}]{Grichener+Soker23}
{Grichener}, A., \& {Soker}, N. 2023, \bibinfo{title}{{Common envelope jets
  supernova with thermonuclear outburst progenitor for the enigmatic supernova
  remnant W49B},} \mnras, 523, 6041, \dodoi{10.1093/mnras/stad1872}

\bibitem[{S. {Hillel} {et~al.}(2023){Hillel}, {Schreier}, \&
  {Soker}}]{Hillel+23}
{Hillel}, S., {Schreier}, R., \& {Soker}, N. 2023, \bibinfo{title}{{Jet-powered
  Turbulence in Common Envelope Evolution},} \apj, 955, 7,
  \dodoi{10.3847/1538-4357/acf19a}

\bibitem[{J.~C. {Houck} \& R.~A. {Chevalier}(1991){Houck} \&
  {Chevalier}}]{Houck+Chevalier91}
{Houck}, J.~C., \& {Chevalier}, R.~A. 1991, \bibinfo{title}{{Steady Spherical
  Hypercritical Accretion onto Neutron Stars},} \apj, 376, 234,
  \dodoi{10.1086/170272}

\bibitem[{N. {Ivanova} {et~al.}(2020){Ivanova}, {Justham}, \&
  {Ricker}}]{Ivanova+20}
{Ivanova}, N., {Justham}, S., \& {Ricker}, P. 2020, {Common Envelope
  Evolution}, \dodoi{10.1088/2514-3433/abb6f0}

\bibitem[{N. {Ivanova} {et~al.}(2013){Ivanova}, {Justham}, {Chen}, {De Marco},
  {Fryer}, {Gaburov}, {Ge}, {Glebbeek}, {Han}, {Li}, {Lu}, {Marsh},
  {Podsiadlowski}, {Potter}, {Soker}, {Taam}, {Tauris}, {van den Heuvel}, \&
  {Webbink}}]{Ivanova+13a}
{Ivanova}, N., {Justham}, S., {Chen}, X., {et~al.} 2013,
  \bibinfo{title}{{Common envelope evolution: where we stand and how we can
  move forward},} \aapr, 21, 59, \dodoi{10.1007/s00159-013-0059-2}

\bibitem[{M.~R. {Krumholz} {et~al.}(2004){Krumholz}, {McKee}, \&
  {Klein}}]{Krumholz+04}
{Krumholz}, M.~R., {McKee}, C.~F., \& {Klein}, R.~I. 2004,
  \bibinfo{title}{{Embedding Lagrangian Sink Particles in Eulerian Grids},}
  \apj, 611, 399, \dodoi{10.1086/421935}

\bibitem[{D. {L{\'o}pez-C{\'a}mara} {et~al.}(2019){L{\'o}pez-C{\'a}mara}, {De
  Colle}, \& {Moreno M{\'e}ndez}}]{Lopez-camara+19}
{L{\'o}pez-C{\'a}mara}, D., {De Colle}, F., \& {Moreno M{\'e}ndez}, E. 2019,
  \bibinfo{title}{{Self-regulating jets during the common-envelope phase},}
  \mnras, 482, 3646, \dodoi{10.1093/mnras/sty2959}

\bibitem[{D. {L{\'o}pez-C{\'a}mara} {et~al.}(2022){L{\'o}pez-C{\'a}mara}, {De
  Colle}, {Moreno M{\'e}ndez}, {Shiber}, \& {Iaconi}}]{Lopez-camara+22}
{L{\'o}pez-C{\'a}mara}, D., {De Colle}, F., {Moreno M{\'e}ndez}, E., {Shiber},
  S., \& {Iaconi}, R. 2022, \bibinfo{title}{{Jets in common envelopes: a
  low-mass main-sequence star in a red giant},} \mnras, 513, 3634,
  \dodoi{10.1093/mnras/stac932}

\bibitem[{D. {L{\'o}pez-C{\'a}mara} {et~al.}(2020){L{\'o}pez-C{\'a}mara},
  {Moreno M{\'e}ndez}, \& {De Colle}}]{Lopez-camara+20}
{L{\'o}pez-C{\'a}mara}, D., {Moreno M{\'e}ndez}, E., \& {De Colle}, F. 2020,
  \bibinfo{title}{{Disc formation and jet inclination effects in common
  envelopes},} \mnras, 497, 2057, \dodoi{10.1093/mnras/staa1983}

\bibitem[{M. {MacLeod} \& E. {Ramirez-Ruiz}(2015){MacLeod} \&
  {Ramirez-Ruiz}}]{Macleod+Ramirez-ruiz15a}
{MacLeod}, M., \& {Ramirez-Ruiz}, E. 2015, \bibinfo{title}{{On the
  Accretion-fed Growth of Neutron Stars during Common Envelope},} \apjl, 798,
  L19, \dodoi{10.1088/2041-8205/798/1/L19}

\bibitem[{E. {Moreno M{\'e}ndez} {et~al.}(2017){Moreno M{\'e}ndez},
  {L{\'o}pez-C{\'a}mara}, \& {De Colle}}]{Morenomendez+17}
{Moreno M{\'e}ndez}, E., {L{\'o}pez-C{\'a}mara}, D., \& {De Colle}, F. 2017,
  \bibinfo{title}{{Dynamics of jets during the common-envelope phase},} \mnras,
  470, 2929, \dodoi{10.1093/mnras/stx1385}

\bibitem[{S.~T. {Ohlmann} {et~al.}(2017){Ohlmann}, {R{\"o}pke}, {Pakmor}, \&
  {Springel}}]{Ohlmann+17}
{Ohlmann}, S.~T., {R{\"o}pke}, F.~K., {Pakmor}, R., \& {Springel}, V. 2017,
  \bibinfo{title}{{Constructing stable 3D hydrodynamical models of giant
  stars},} \aap, 599, A5, \dodoi{10.1051/0004-6361/201629692}

\bibitem[{P.~A. {Ondratschek} {et~al.}(2022){Ondratschek}, {R{\"o}pke},
  {Schneider}, {Fendt}, {Sand}, {Ohlmann}, {Pakmor}, \&
  {Springel}}]{Ondratschek+22}
{Ondratschek}, P.~A., {R{\"o}pke}, F.~K., {Schneider}, F. R.~N., {et~al.} 2022,
  \bibinfo{title}{{Bipolar planetary nebulae from common-envelope evolution of
  binary stars},} \aap, 660, L8, \dodoi{10.1051/0004-6361/202142478}

\bibitem[{B. {Paczynski}(1976){Paczynski}}]{Paczynski76}
{Paczynski}, B. 1976, \bibinfo{title}{{Common Envelope Binaries},} in IAU
  Symposium, Vol.~73, Structure and Evolution of Close Binary Systems, ed.
  P.~{Eggleton}, S.~{Mitton}, \& J.~{Whelan}, 75

\bibitem[{O. {Papish} \& N. {Soker}(2011){Papish} \& {Soker}}]{Papish+Soker11}
{Papish}, O., \& {Soker}, N. 2011, \bibinfo{title}{{Exploding core collapse
  supernovae with jittering jets},} \mnras, 416, 1697,
  \dodoi{10.1111/j.1365-2966.2011.18671.x}

\bibitem[{B. {Paxton} {et~al.}(2011){Paxton}, {Bildsten}, {Dotter}, {Herwig},
  {Lesaffre}, \& {Timmes}}]{Paxton+11}
{Paxton}, B., {Bildsten}, L., {Dotter}, A., {et~al.} 2011,
  \bibinfo{title}{{Modules for Experiments in Stellar Astrophysics (MESA)},}
  \apjs, 192, 3, \dodoi{10.1088/0067-0049/192/1/3}

\bibitem[{B. {Paxton} {et~al.}(2013){Paxton}, {Cantiello}, {Arras}, {Bildsten},
  {Brown}, {Dotter}, {Mankovich}, {Montgomery}, {Stello}, {Timmes}, \&
  {Townsend}}]{Paxton+13}
{Paxton}, B., {Cantiello}, M., {Arras}, P., {et~al.} 2013,
  \bibinfo{title}{{Modules for Experiments in Stellar Astrophysics (MESA):
  Planets, Oscillations, Rotation, and Massive Stars},} \apjs, 208, 4,
  \dodoi{10.1088/0067-0049/208/1/4}

\bibitem[{B. {Paxton} {et~al.}(2015){Paxton}, {Marchant}, {Schwab}, {Bauer},
  {Bildsten}, {Cantiello}, {Dessart}, {Farmer}, {Hu}, {Langer}, {Townsend},
  {Townsley}, \& {Timmes}}]{Paxton+15}
{Paxton}, B., {Marchant}, P., {Schwab}, J., {et~al.} 2015,
  \bibinfo{title}{{Modules for Experiments in Stellar Astrophysics (MESA):
  Binaries, Pulsations, and Explosions},} \apjs, 220, 15,
  \dodoi{10.1088/0067-0049/220/1/15}

\bibitem[{B. {Paxton} {et~al.}(2019){Paxton}, {Smolec}, {Schwab}, {Gautschy},
  {Bildsten}, {Cantiello}, {Dotter}, {Farmer}, {Goldberg}, {Jermyn}, {Kanbur},
  {Marchant}, {Thoul}, {Townsend}, {Wolf}, {Zhang}, \& {Timmes}}]{Paxton+19}
{Paxton}, B., {Smolec}, R., {Schwab}, J., {et~al.} 2019,
  \bibinfo{title}{{Modules for Experiments in Stellar Astrophysics (MESA):
  Pulsating Variable Stars, Rotation, Convective Boundaries, and Energy
  Conservation},} \apjs, 243, 10, \dodoi{10.3847/1538-4365/ab2241}

\bibitem[{R. {Pettibone} \& E.~G. {Blackman}(2009){Pettibone} \&
  {Blackman}}]{Pettibone+2009}
{Pettibone}, R., \& {Blackman}, E.~G. 2009, \bibinfo{title}{{Stochastic wobble
  of accretion discs and jets from turbulent rocket torques},} \mnras, 396,
  1783, \dodoi{10.1111/j.1365-2966.2009.14863.x}

\bibitem[{L.~J. {Prust} \& P. {Chang}(2019){Prust} \& {Chang}}]{Prust+Chang19}
{Prust}, L.~J., \& {Chang}, P. 2019, \bibinfo{title}{{Common envelope evolution
  on a moving mesh},} \mnras, 486, 5809, \dodoi{10.1093/mnras/stz1219}

\bibitem[{P.~M. {Ricker} \& R.~E. {Taam}(2008){Ricker} \&
  {Taam}}]{Ricker+Taam08}
{Ricker}, P.~M., \& {Taam}, R.~E. 2008, \bibinfo{title}{{The Interaction of
  Stellar Objects within a Common Envelope},} \apjl, 672, L41,
  \dodoi{10.1086/526343}

\bibitem[{P.~M. {Ricker} \& R.~E. {Taam}(2012){Ricker} \&
  {Taam}}]{Ricker+Taam12}
{Ricker}, P.~M., \& {Taam}, R.~E. 2012, \bibinfo{title}{{An AMR Study of the
  Common-envelope Phase of Binary Evolution},} \apj, 746, 74,
  \dodoi{10.1088/0004-637X/746/1/74}

\bibitem[{F.~K. {Roepke} \& O. {De Marco}(2022){Roepke} \& {De
  Marco}}]{Roepke+Demarco22}
{Roepke}, F.~K., \& {De Marco}, O. 2022, \bibinfo{title}{{Simulations of
  common-envelope evolution in binary stellar systems: physical models and
  numerical techniques},} arXiv e-prints, arXiv:2212.07308,
  \dodoi{10.48550/arXiv.2212.07308}

\bibitem[{R. {Schreier} {et~al.}(2023){Schreier}, {Hillel}, \&
  {Soker}}]{Schreier+23}
{Schreier}, R., {Hillel}, S., \& {Soker}, N. 2023, \bibinfo{title}{{Simulating
  the deposition of angular momentum by jets in common envelope evolution},}
  \mnras, 520, 4182, \dodoi{10.1093/mnras/stad360}

\bibitem[{N. {Soker}(2016){Soker}}]{Soker16}
{Soker}, N. 2016, \bibinfo{title}{{The jet feedback mechanism (JFM) in stars,
  galaxies and clusters},} \nar, 75, 1, \dodoi{10.1016/j.newar.2016.08.002}

\bibitem[{N. {Soker}(2017){Soker}}]{Soker17b}
{Soker}, N. 2017, \bibinfo{title}{{Energizing the last phase of common-envelope
  removal},} \mnras, 471, 4839, \dodoi{10.1093/mnras/stx1978}

\bibitem[{N. {Soker}(2021){Soker}}]{Soker21}
{Soker}, N. 2021, \bibinfo{title}{{Double common envelope jets supernovae
  (CEJSNe) by triple-star systems},} \mnras, 504, 5967,
  \dodoi{10.1093/mnras/stab1275}

\bibitem[{Y. {Zou} {et~al.}(2022){Zou}, {Chamandy}, {Carroll-Nellenback},
  {Blackman}, \& {Frank}}]{Zou+22}
{Zou}, Y., {Chamandy}, L., {Carroll-Nellenback}, J., {Blackman}, E.~G., \&
  {Frank}, A. 2022, \bibinfo{title}{{Jets from main sequence and white dwarf
  companions during common envelope evolution},} \mnras, 514, 3041,
  \dodoi{10.1093/mnras/stac1529}

\end{thebibliography}
\bibliographystyle{aasjournalv7}



\end{document}